%% file: ms.tex
\documentclass[12pt]{emulateapj}

\usepackage{natbib, graphicx, amsmath}

\newcommand{\name}{J1009+0713}
\newcommand{\kms}{km s$^{-1}$}
\newcommand{\tnote}{\tablenotemark}
\newcommand{\6}{\ion{O}{6}}

\shorttitle{Multiphase Galaxy Halo Gas Toward J1009+0713}
\shortauthors{Tumlinson et al.}

\begin{document}

\title{Multiphase Gas in Galaxy Halos: The \ion{O}{6} Lyman-Limit System Toward J1009+0713\altaffilmark{1}}

\author{J. Tumlinson\altaffilmark{2}, 
J. K. Werk\altaffilmark{3,4}, 
C. Thom\altaffilmark{2,4}, 
J. D. Meiring\altaffilmark{5},
J. X. Prochaska\altaffilmark{3,4}, 
T. M. Tripp\altaffilmark{5}, 
J. M. O'Meara\altaffilmark{6}, 
M. Okrochkov\altaffilmark{2},
K. R. Sembach\altaffilmark{2}}
\altaffiltext{1}{Based on observations made with the NASA/ESA Hubble Space Telescope, obtained at the Space Telescope Science Institute, which is operated by the Association of Universities for Research in Astronomy, Inc., under NASA contract NAS 5-26555. These observations are associated with program GO11598.}
\altaffiltext{2}{Space Telescope Science Institute, 3700 San Martin Drive, Baltimore, MD}
\altaffiltext{3}{University of California Observatories-Lick Observatory, UC Santa Cruz, CA}
\altaffiltext{4}{Visiting Astronomer, W.M. Keck Telescope.The Keck Observatory is a joint facility of the University of California and the California Institute of Technology.}
\altaffiltext{5}{Department of Astronomy, University of Massachusetts, Amherst, MA} 
\altaffiltext{6}{Department of Chemistry and Physics, Saint Michael's College, One Winooski Park, Colchester, VT}

\begin{abstract}
We have  serendipitously detected a strong \ion{O}{6}-bearing Lyman limit system at $z_{abs} = 0.3558$ toward the QSO \name\ ($z_{em} = 0.456$) in our survey of low-redshift galaxy halos with the Hubble Space Telescope's {\it Cosmic Origins Spectrograph}. Its total rest-frame equivalent width of $W_r = 835 \pm 49$ m\AA\ and column density of $\log N$(\6) = 15.0 are the highest for an intervening absorber yet detected in any low-redshift QSO sightline, with absorption spanning at least four major kinematic component groups over 400 km s$^{-1}$ in its rest frame.  HST/WFC3 images of the galaxy field show that the absorber is associated with two galaxies lying at 14 and 46 kpc from the QSO line of sight. The absorber is kinematically complex and there are no less than nine individual \ion{Mg}{2}components spanning 200 km s$^{-1}$ in our Keck/HIRES optical data. The bulk of the absorbing gas traced by \ion{H}{1} resides in two strong, blended component groups that possess a total $\log N$(\ion{H}{1}) $\simeq 18 - 18.8$, but most of the \ion{O}{6} is associated with two outlying components with $\log N$(\ion{H}{1}) = 14.8 and 16.5. The ion ratios and column densities of  C, N, O, Mg, Si, S, and Fe, except the \ion{O}{6}, can be accommodated into a simple photoionization model in which diffuse, low-metallicity halo gas is exposed to a photoionizing field from stars in the nearby galaxies that propagates into the halo at 10\% efficiency. In this model, the clouds have neutral fractions of $\sim 1 - 10$\% and thus total hydrogen column densities of $\log N$(H)  $\simeq 19.5$. Direct measurement of the gas metallicity is precluded by saturation of the main components of the \ion{H}{1}, but we constrain the metallicity firmly within the range 0.1 - 1 $Z_{\odot}$, and photoionization modeling indirectly indicates a subsolar metallicity of 0.05 - 0.5 $Z_{\odot}$. This highly ionized, multiphase, possibly low-metallicity halo gas resembles gas with similar properties in the Milky Way halo and other low-redshift LLS, suggesting that at least some other galaxies have their star formation fueled by metal-poor gas accreting from the intergalactic medium and ionized by the stars in the host galaxy. As observed in the Milky Way high-velocity clouds, the strong detected \ion{O}{6} is not consistent with the photoionization scenario but is consistent with general picture in which \ion{O}{6} arises in interface material surrounding the photoionized clouds or in a hotter, diffuse component of the halo. The appearance of strong \ion{O}{6} and nine \ion{Mg}{2} components in this system, and our review of similar systems in the literature, offer some support to this ``interface" picture of high-velocity \ion{O}{6}: the total strength of the \ion{O}{6} shows a positive correlation with the number of detected components in the low-ionization gas. 
\end{abstract}



\section{Introduction}

\begin{figure*}[!t]
\plotone{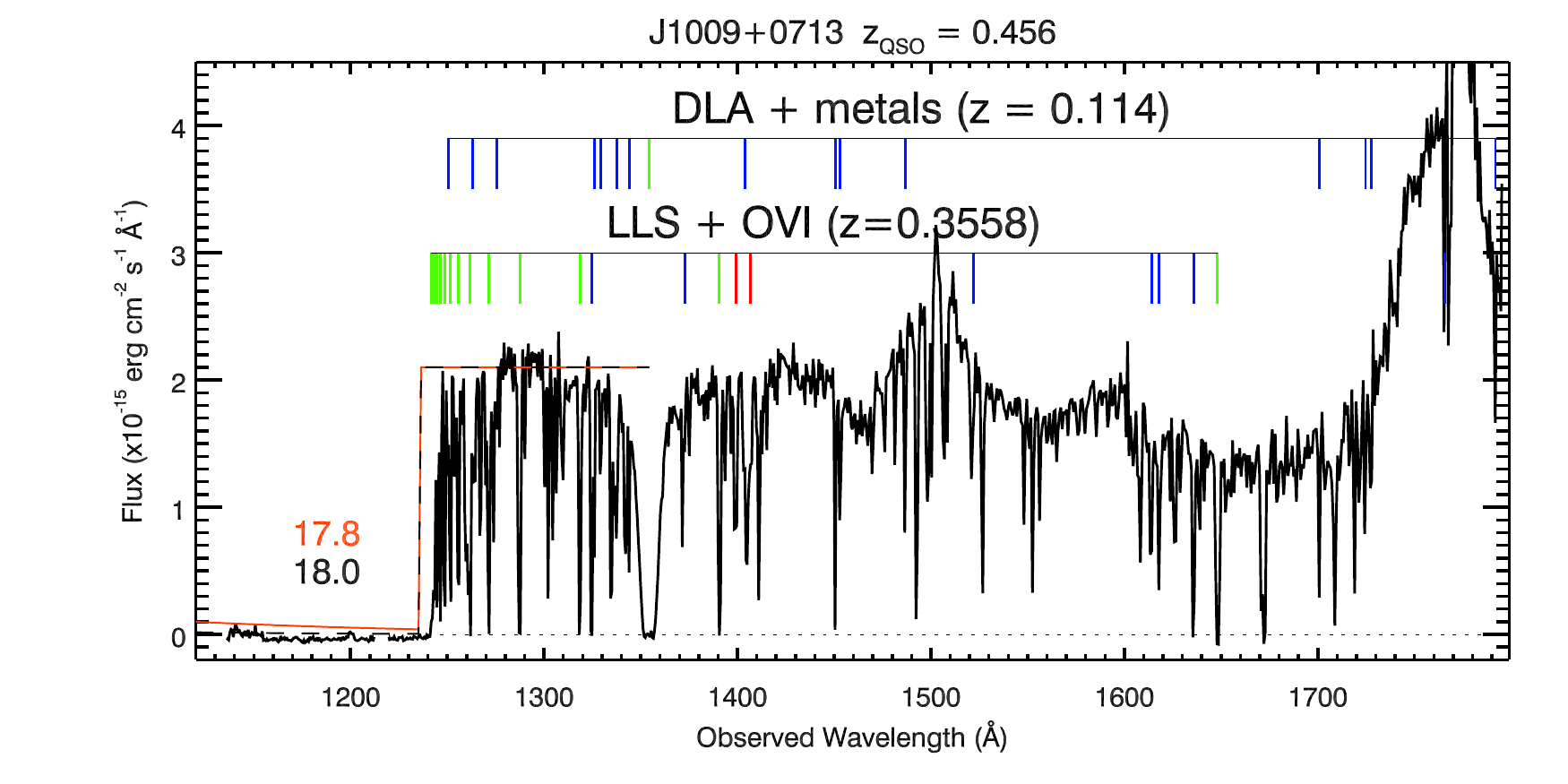}
\caption{The co-added COS data, plotted versus observed wavelength, and binned by 3 raw COS pixels to optimal sampling of 2 bins per resolution element. The Lyman limit system analyzed in this paper and the DLA from \cite{Meiring:11:1} are marked at the positions of Lyman series lines (green), metal lines (blue), and \ion{O}{6} (red). The Lyman limit of the present system is clearly visible as a complete absorption of the QSO spectrum at $\sim 1240$ \AA. Geocoronal Ly$\alpha$ emission has been excised near $1216$ \AA. The solid and dashed curves mark the opacity of the Lyman limit system using total column densities of $\log N$(\ion{H}{1}) = 17.8 and 18.0, respectively, as described in \S~3.5. \label{wholespecfig}}
\vspace{0.05in}
\end{figure*}

A full understanding of how galaxies acquire their gas from the IGM and return it there in the form of chemical and kinematic feedback will likely form an important part of any complete picture of galaxy formation. Motivated by questions of how galaxies obtain their observed stellar masses and morphology, theorists have developed a picture in which gas enters dark matter halos and galaxies by either cold flows along filaments \citep{Keres:05:2a}, accretion of hot material that has passed through a shock on its way in from the IGM \citep{Dekel:06:2}, or in some complex, multiphase mixture of the two \citep{Maller:04:694}. Mass loss by feedback is driven in ``superwind'' ouflows fueled by many correlated supernovae, by radiation pressure on dust \citep{Murray:05:569}, or in AGN-triggered flows in those galaxies that possess active nuclei. Tidal or ram-pressure stripping of gas from satellite galaxies during gravitational encounters may also provide a gas supply to larger galaxies, or at least to their halos. All these gas processes may bear on such observational features of galaxies as their stellar masses, morphologies, and colors, but to be effective they must act across the 100 kpc scales of galaxy halos. Unfortunately the proposed accretion, feedback, and stripping processes are difficult to test directly because diffuse, ionized gas in the immediate vicinity of galaxies is difficult to detect. We therefore have an incomplete empirical picture of how accreting gas is distributed around galaxies of all types, what its temperature, density, and metallicity configurations look like, and how these features influence or are influenced by the host galaxy. 

The classic quasar absorption-line technique provides a method for studying intergalactic and circumgalactic medium gas, even to very low density and metallicity. This technique has been used effectively to probe diffuse ionized gas in galaxy halos,  over long pathlengths through the IGM, and throughout the halo of the Milky Way. The IGM samples reveal the statistical correlations of galaxies with intervening absorbers, and in some cases show physical properties of gas that is well within the virial radius. However, most such information comes from {\it post-facto} galaxy surveys that do not uniformly sample the range of galaxy properties that may be related to the gas properties. While extensive observational information exists about the quantity and physical state of multiphase high-velocity cloud (HVC) gas in the Milky Way halo \citep{Sembach:03:165, Fox:05:332, Collins:05:196, Shull:09:754}, and gas stripping from dwarf galaxies is clearly demonstrated in the case of the Magellanic Clouds and Stream, it is not known how typical the Milky Way is in this regard. 

To address the problem of how halo gas in other galaxies is distributed and how it compares to the halo gas of the Milky Way, we have begun an effort to systematically survey the gaseous halos of low-redshift galaxies using the Cosmic Origins Spectrograph (COS) aboard the {\it Hubble Space Telescope}. Our survey chose galaxies with sightlines to UV-bright QSOs passing through their halos, with redshifts tuned to place the $\lambda\lambda$1032,1038 doublet of \ion{O}{6} near the peak of COS sensitivity at $1250-1350$ \AA. This survey design allows us access to all the key far-UV ionization diagnostics from which gas budgets, metallicities, and kinematics can be inferred. As part of this survey we are also obtaining galaxy spectra and high-resolution optical spectra of the QSOs. The main results of this survey will be published elsewhere; here we report on a serendipitous discovery from this program that nevertheless addresses its main goal: to examine the gaseous fuel and/or waste residing in galactic halos. 

This paper reports new HST/COS and HST/WFC3 and Keck HIRES and LRIS data for the sightline to \objectname{J1009+0713}. This sightline exhibits a strong Lyman-limit system (LLS) bearing a wide range of ionic absorption from multiple ionization stages and in several distinct kinematic components. The system was discovered serendipitously in the data on this quasar obtained as part of our survey; the absorption-line system associated with the targeted galaxy along this sightline will be published separately. This paper first describes the data collection and analysis for the HST/COS and Keck/HIRES data on the sightline to the QSO, and on our HST/WFC3 and Keck/LRIS data on the galaxies in the field (\S~2). We then report the details of our line identification and analysis (\S~\ref{lls-section}), and on the analysis of the galaxy images and spectra (\S~\ref{gal-section}). Section 5 describes physical models for the absorber. In Section~\ref{interp-section} we describe our general results and their significance for the larger picture of gas in galaxy halos. We adopt the WMAP7 cosmological parameters $H_0 = 100h= 71$ \kms\ Mpc$^{-1}$, $\Omega _{\Lambda} = 0.734$, and $ \Omega _{m} h^2= 0.1334$ \citep{Komatsu:10:4538}. 

\section{Observations}

\begin{figure*}
\begin{center} 
\includegraphics[width=5.0 in]{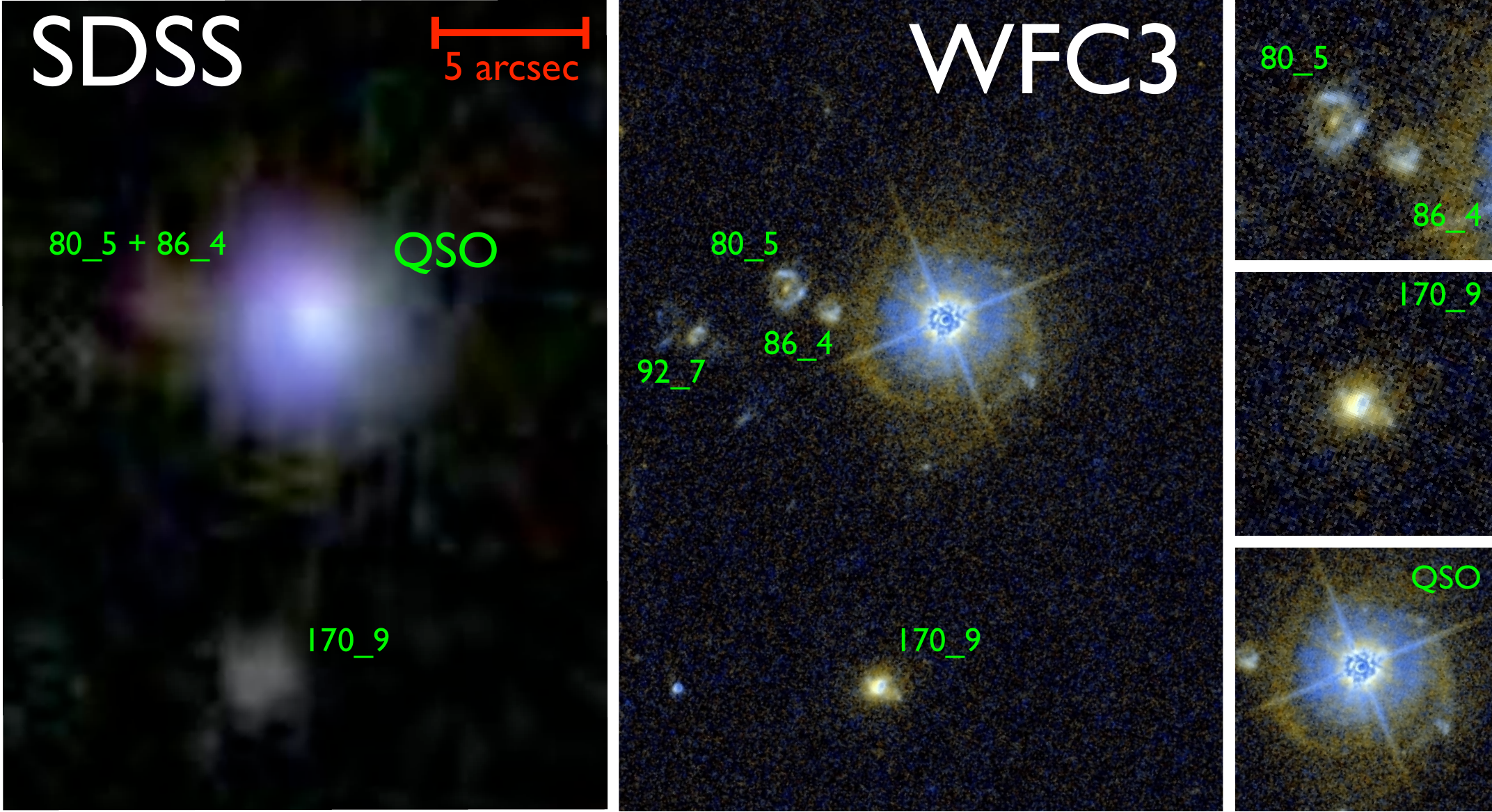}
\caption{The galaxy field, aligned with north at top and east to the left. The SDSS Skyserver image and our WFC3 image are shown for contrast.  The larger panels are approximately 25 arcsec across. Galaxies 170\_9 and 86\_4 are at $z = 0.356$, the redshift of the LLS. At the redshift of the system, the $5"$ range bar subtends $25h^{-1}$ kpc. Precise impact parameters for the galaxies are given in Table~2. The insets show zoomed images of the three galaxies of interest and the QSO itself. \label{fieldfig}}
\end{center} 
\vspace{0.05in}
\end{figure*}

\subsection{COS Data} 

Data on the SDSSJ100902.06+071343.8 sightline (hereafter J1009+0713) were obtained by COS over 3 orbits on 29-30 Mar 2010 as Visit 13 in Program GO11598. Two exposures were obtained with the FUV G130M grating at central wavelength settings 1291 and 1309, with total exposure times of 1497 and 2191 sec, respectively. Two exposures were obtained with the FUV G160M grating at central wavelengths settings 1577 and 1600, with exposure times of 2002 and 2007 sec, respectively. All exposures were taken in TIME-TAG readout mode with contemporaneous wavelength stim pulses and the default FP-POS position. The data were processed with the standard CALCOS data pipeline (v2.12) on retrieval from the archive. More details on the performance of COS can be found in the COS Instrument Handbook \citep{Dixon:10:202} and at the STScI website. 

We began our analysis with the {\tt x1d} files provided by CALCOS. These extracted 1D spectra were then coadded to combine the multiple exposures for each grating, and in turn the two gratings, into a single spectrum. To align the four exposures as closely as possible in wavelength space, small ($\lesssim10$ pixel) shifts were derived for each exposure individually from the profiles of strong Galactic ISM absorption lines.  This coaddition was performed on the ``gross counts'' vector from the CALCOS {\tt x1d} files. The co-addition process sums counts from separate exposures for each pixel, keeping track of the effective exposure time for each pixel correctly. That is, if a given pixel is ignored in a particular exposure because of data quality flags, that exposure time is not summed for that pixel. Thus we are effectively co-adding {\em count rates} rather than counts. Co-addition in count space allows for a simple derivation of the correct error vector in the Poisson limit of low total counts \citep{1986ApJ...303..336G}. The final coadded spectrum possesses a signal-to-noise ratio $S/N = 7 - 15$ per resolution element. The resolving power of COS is approximately $R = \lambda / \Delta \lambda = 16-18,000$ over 1140 - 1750 \AA, and the resolution element is sampled by approximately six (analog) pixels in the 1D extraction. We binned our final, coadded spectrum into 3-pixel bins for analysis. This binning leaves approximately 2 bins per resolution element, close to optimal sampling. 

We note that the adopted version of CALCOS did not apply a correction for fixed pattern noise, but that the coaddition of exposures taken with different central wavelength settings mitigates the regular shadowing pattern of the detector quantum-efficiency (DQE) grid wires lying above the face of the COS micro-channel plate detectors. During the coaddition of the data we applied a correction for these grid wire shadows provided to us by the COS instrument team. This correction removes the grid-wire features, which otherwise result in $\sim 15$\% less counts recorded over a few pixels and can mimic absorption lines. 

The final, reduced COS data appear in Figure~\ref{wholespecfig}. The spectrum reveals at least two absorption line systems of significance along this sightline. First, we discovered the $z_{abs} = 0.356$ Lyman-limit system (LLS) that is the main focus of this paper. This system presents an obvious Lyman-limit break at 1240 \AA\ and a host of other multiphase ionization stages that will be analyzed below. A damped-Lyman-$\alpha$ system (DLA) was discovered at $z_{abs} = 0.114$ and has been analyzed by \cite{Meiring:11:1}. Both of these latter systems were serendipitous discoveries along a sightline that was selected to pass through the halo of an unrelated galaxy at $z < 0.25$, and we have no reason to suspect that the selection of the QSO itself or the targeted galaxy introduced a bias in favor of these two extraordinary intervening systems. The properties of the targeted galaxy and its associated absorber will be presented as part of the main survey. 


\subsection{WFC3 Data}

Based on the SDSS images, we identified two galaxies that could be associated with the LLS at $z = 0.356$ and/or the DLA at $z = 0.114$. These are labeled ``80\_5+86\_4'' and ``170\_9'' in Figure~\ref{fieldfig} (left panel)\footnote{We label our galaxies with a position angle and angular separation in arcseconds with respect to the target QSO; thus galaxy 170\_9 has position angle of 170$^{\circ}$ degrees (N through E) and is separated by 9$"$ from J1009+0713.}. The indistinct profile near the QSO is partially hidden from view by the ground-based PSF of J1009+0713 in the SDSS image. Any attempt to relate the LLS to the galaxy properties, or even to obtain an accurate measurement of the impact parameter of this galaxy to the QSO sightline, was significantly hindered by this source confusion. In an attempt to conclusively identify the LLS-associated galaxies and to possibly discover a DLA host galaxy close to the sightline (fully within the SDSS PSF of J1009+0713), we obtained an image of this field with the Wide Field Camera 3 aboard {\it HST}.

The WFC3/UVIS data on this field were obtained on 24 June 2010 as Visit 44 in program GO11598. We obtained images in two broadband filters, F390W and F625W. The QSO was placed at the ``UVIS1'' aperture position, and the WFC3-UVIS-MOS-DITH-LINE dither pattern was used with an exposure time of 376 sec in F625W and 395 sec in F390W. This pattern resulted in total exposure times of 2256 sec and 2370 sec in F625W and F390W, respectively. The six individual exposures in each filter were first processed by the default CALWFC3 pipeline. Each exposure then had the QSO subtracted off using a TinyTim model of the telescope PSF before the six exposures were added together using MultiDrizzle. More details about the processing of the WFC3 image can be found in the paper by \cite{Meiring:11:1}. 

Even without coaddition or QSO PSF subtraction, the WFC3 image revealed that the candidate galaxy nearest the QSO in the SDSS image resolves into two galaxies, labeled 80\_5 and 86\_4 in Figure~\ref{fieldfig}, that are separated by only $\sim 1''$. The image itself does not tell us whether one or more of these galaxies is associated with the DLA or LLS. We have attempted to constrain their redshifts separately with followup LRIS data, as described in the next section. 

\subsection{Keck LRIS}

After the COS data revealed two unexpected and interesting absorbers in this sightline, we re-examined the imaging data from which we had selected the target QSO. To determine which of these galaxies, if any, were associated with our newly detected absorbers, we obtained 1\arcsec-wide longslit spectra of these two galaxy candidates with the Keck/Low-Resolution Imaging Spectrometer (LRIS) on 5 April 2010. These LRIS data were taken using the D560 dichroic with the 600/4000 l/mm grism (blue side) and 600/7500 l/mm grating (red side) which gives a spectral coverage between 3000 and 5500 \AA\ (blue side), and 5600 to 8200 \AA\ (red side). On the blue side, binning the data 2 $\times$ 2 resulted in a dispersion of 1.2 \AA\ per pixel and a FWHM resolution of $\sim$280 km/s. On the red side, the data were binned 1 $\times$ 2, resulting in a dispersion of 2.3 \AA\ per pixel and a FWHM resolution of  $\sim$200 km/s. Exposure times were 800s in the blue and 2 $\times$ 360 s in the red, which resulted in signal-to-noise ratios of at least 3 per pixel for strong nebular emission lines in the galaxy spectra. 

\begin{figure}
\epsscale{1.1}
\plotone{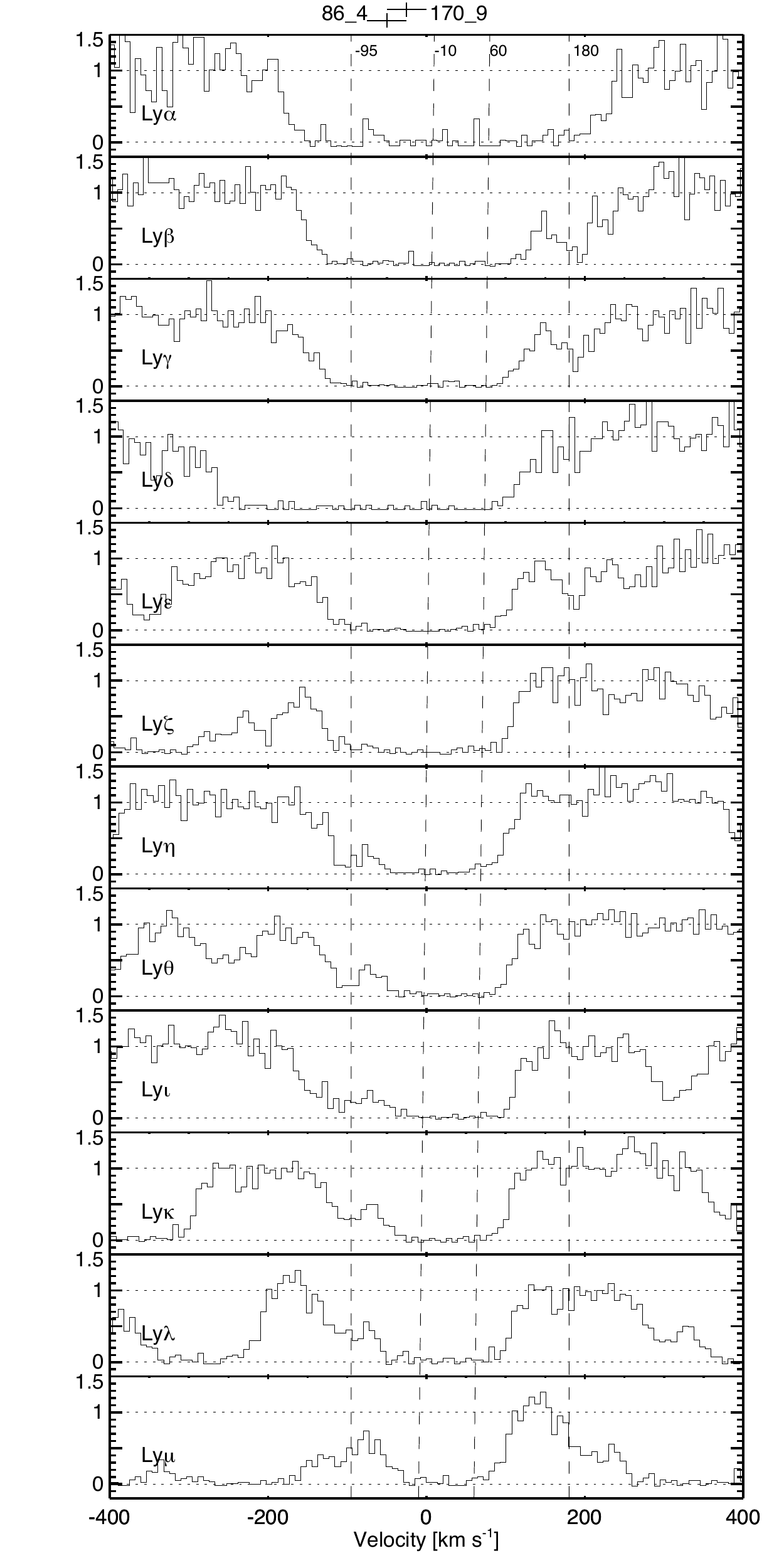}
\caption{Lyman series lines for the $z = 0.355$ LLS. The velocity zero point is set to $z = 0.3558$. The points with error bars at the top mark the measured velocities of the galaxies with 25 \kms\ uncertainty. \label{s0.355_ly}}
\end{figure}

\begin{figure*}
\epsscale{1}
\plotone{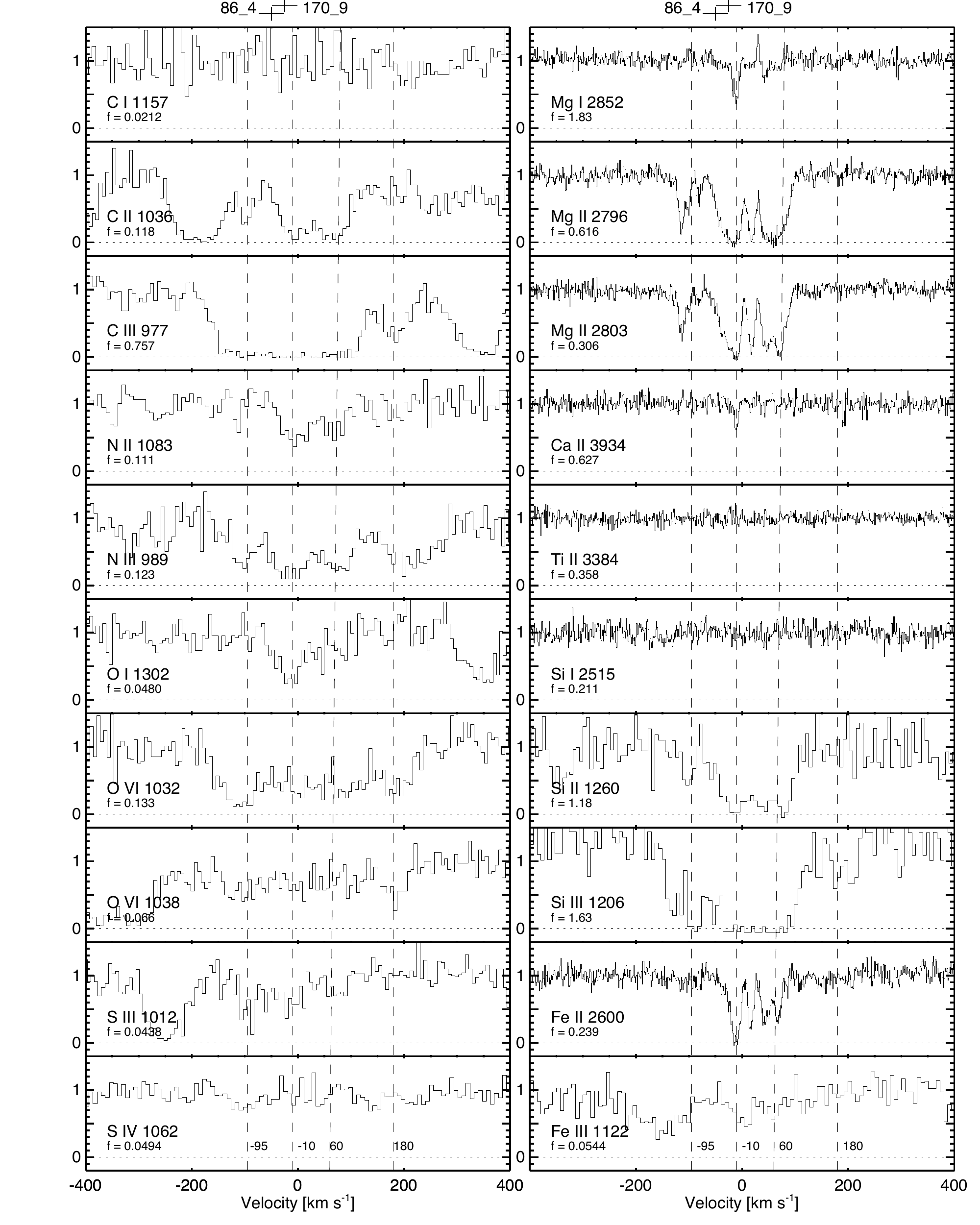}
\caption{Metal absorption lines from the $z = 0.3558$ LLS. Vertical dashed lines mark the centroid velocities of components A - D. Transitions are ordered by atomic element and, within that, by increasing ionization potential. The points with error bars at the top mark the measured velocities of the two associated galaxies with 25 \kms\ uncertainty.  \label{metal_stack}}
\end{figure*}

Data reduction and calibration were carried out using the LowRedux\footnote{http://www.ucolick.org/$\sim$xavier/LowRedux/index.html} IDL software package, which includes flat fielding to correct for pixel-to-pixel response variations and larger scale illumination variations, wavelength calibration, sky subtraction, and flux calibration using the spectrophotometric standard star G191B2B.  Precise and accurate systemic redshifts were obtained for both galaxies using a modified version of the SDSS IDL code ``zfind'', which works by fitting smoothed template SDSS eigenspectra to the galaxy emission-line spectra on both red and blue sides. The resultant weighted-mean redshift for 170\_9 is $z = 0.355687 \pm 0.00001$. The long slit used to obtain this first set of Keck/LRIS observations was oriented to run from galaxy 170\_9 through the midpoint of the faint profile seen for 80\_5+86\_4 in the SDSS image (the WFC3 image was not yet available). The slit was 1$''$ wide, so the recorded spectrum most likely includes contributions from both galaxies 80\_5 and 86\_4. For this spectrum, we obtain $z = 0.355574 \pm 0.00002$, with error bars calculated from statistical noise only. To account for systematic uncertainties in the absolute wavelength calibration and instrument flexure during the LRIS exposures, we adopt a larger 25 \kms\ error on the galaxy redshifts. The redshifts of the two galaxies are marked with ticks and $25$ \kms\ errors at the top of Figures~\ref{s0.355_ly}, \ref{metal_stack}, and \ref{kinplotfig} after translation into the rest frame of the absorber ($z_{abs} = 0.3558$). The galaxies appear approximately 25 \kms\ apart. Neither appears to line up exactly with any of the strongest absorption components, but their coincidence is enough to give us strong evidence of their kinematic association. 

Additional LRIS exposures of this field were obtained with LRIS in January 2011 in an attempt to separately constrain the redshifts of 80\_5 and 86\_4, which were blended together in the first set of LRIS exposures. LRIS was configured using the 400/3400 grism on the blue side, the 600/7500 grating on the red side, and the d650 dichroic.  This setup provided full coverage of wavelengths $3000 < \lambda < 8700$ \AA.  Three exposures, each lasting 600 seconds, were obtained in $\sim 1$ arcsecond seeing.  Each exposure had a different position angle, with the goal of obtaining spectra of 80\_5, 86\_4, and other sources, while minimizing contaminating flux from the QSO.  This observation confirmed the detection of the [\ion{O}{3}] and H$\beta$ lines at $z = 0.355$ for 86\_4. With the separation of 80\_5 and 86\_4 measured from the WFC3 image, we were able to marginally separate the spectral traces of 80\_5 and 86\_4. We do not find the same pattern of emission lines at the position of 80\_5, so we conclude that this galaxy is not at the same redshift as 86\_4. The best solution for the weak emission line detections at this position gives a solution of $z = 0.879$.  We therefore disregard this galaxy in our subsequent analysis. 

\subsection{Keck HIRES}

Using the Keck I  High Resolution Echelle Spectrometer with the blue collimator (HIRESb), we obtained a high dispersion spectrum of the J1009+0713 on 26 March 2010.  These data were taken through the C1 decker, resulting in a FWHM resolution of $\approx 6$ \kms. The grating angles were set to provide wavelength coverage from $3050-5870$ \AA, with two small gaps related to the separation of the three CCD mosaic.  We integrated for two 1800s exposures under good conditions.  The data were reduced with the HIRedux pipeline\footnote{http://www.ucolick.org/$\sim$xavier/HIRedux/index.html} to flatfield, sky subtract, wavelength calibrate, and extract the spectra.  The data were optimally coadded and normalized to unit flux by fitting a series of Chebyshev polynomials.  The final spectrum has a S/N of 15 per 1.3 \kms\ pixel redwards of 3800\AA\ decreasing to S/N=5 at 3200 \AA.

\section{Absorption Line Identification and Analysis}
\label{lls-section}

\subsection{General Approach}

To identify and fit absorption lines in the spectrum of J1009+0713 we have adopted the vacuum wavelengths and atomic data from the compilation by \cite{Morton:03:205}. Line measurements appear in Table~\ref{comp-list}. Most column densities were obtained by direct integration over the line profile, except in the case of the optical lines,  some saturated UV lines, and the Lyman series lines in the outlying components, where profile fits were used. 

The line profile fits use a custom Voigt-profile fitting software that attempts to optimize a model of the line column densities, doppler $b$ parameters, and velocities for an arbitrary input set of lines and atomic data, by minimizing the $\chi ^2$. This code was first used to analyze the PG1211+143 sightline by \cite{Tumlinson:05:95}. This software first constructs intrinsic line profiles and then convolves these with the appropriate non-Gaussian COS line-spread function (LSF) as described by \cite{Ghavamian:09:1} and available on the STScI COS website. These tabulated LSFs are specified for 50 \AA\ intervals through the range of the G130M and G160M gratings. For line fits we adopt the nearest LSF grid point based on the observed wavelength of the lines being fitted.

\subsection{The Lyman Limit System}

The $z = 0.3558$ absorber exhibits a line-black Lyman limit at 1236 \AA\ in the observed frame. This system also appears in at least 14 distinct Lyman series transitions, the first twelve of which appear in Figure~\ref{s0.355_ly}. Inspection of the strong Ly$\alpha$ profile reveals absorption extending over $\sim 400$ \kms. No damping wings are apparent, which limits the total column density to $\log N$(\ion{H}{1}) $\lesssim 19$. At Ly$\beta$ - $\epsilon$, a separate component at $+180$ \kms\ becomes distinct, which we term component D. By Ly$\eta$, a separate component at $-95$ \kms\ is visible and distinct until blending between the Lyman series lines themselves becomes severe at Ly$\mu$. The two innermost components labeled B and C in Figure~\ref{s0.355_ly} never separate in the Lyman series lines and are defined by their distinct profiles in the metal lines, particularly \ion{Mg}{2}, as shown below. Component groups A, B, and C all break further into multiple blended components in the HIRES data. 

We note that these components are defined, wherever possible, by the \ion{H}{1} absorption or by the metal lines, as noted in Table 1. The number of apparent components is larger in the higher resolution optical data covering \ion{Mg}{2} and \ion{Fe}{2}. This finding suggests that there may be further component structure in the \ion{H}{1} and metal lines that we cannot resolve at the R = 18,000 resolution of COS. In what follows, we analyze the major component groups as coherent objects with the proviso that they may in fact consistent of subcomponents below the level of our spectral resolution.

\input t1.tex

\subsection{Metal lines and ionization}

The COS data on the LLS reveal absorption from a wide range of ionization stages, as shown in Figure~\ref{metal_stack}. We detect absorption by C, N, O, Mg, Ca, Si, S, and Fe in the COS and/or HIRES data, and in ionization stages from \ion{Mg}{1} to \ion{O}{6}. However, the overall impression is of an absorption-line system that possesses a high degree of ionization. The species \ion{Mg}{1} and \ion{O}{1} are the only neutral atoms that are convincingly detected; \ion{Ca}{2} is usually detected in mostly neutral gas but is very weak here. Limits are placed on the neutrals of C, N, Si, and Fe. By contrast, strong absorption is detected in \ion{C}{3}, \ion{N}{3}, \ion{Si}{3}, and \ion{Fe}{3}.  Our COS data do not cover \ion{C}{4} or \ion{Si}{4}, but \ion{N}{5} $\lambda\lambda 1238,1242$, typically weak, is not detected in the S/N  $\sim 5$ data at that wavelength. These detections and limits, apart from any detailed analysis, point to gas that is substantially ionized in all components. 

\subsection{The O VI Absorption}

\begin{figure}[!t]
\epsscale{1}
\plotone{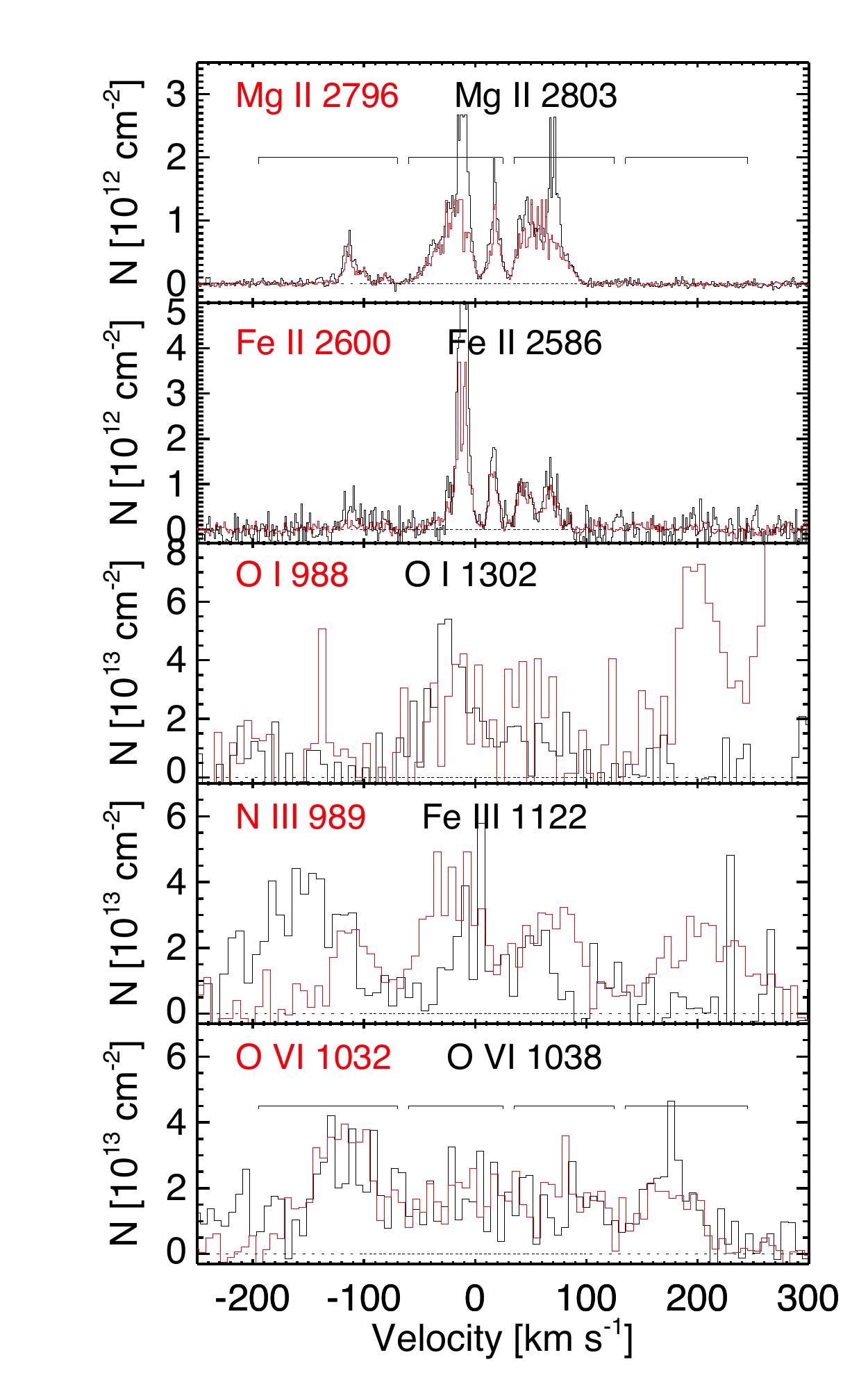}
\caption{Kinematic structure in apparent column densities \citep{Savage:91:245} for low ions and \ion{O}{6}.  Major components A, B, and C are broken down further according to their appearance in the Mg II and Fe II data obtained by HIRES. Note the wide range of apparent column density ratios between the low ions and \ion{O}{6}, particularly in components A and D. Flat tops arise from the truncation of the flux at 0.01 in the normalized data to avoid unphysical values in the optical depth for saturated lines. \label{kinplotfig}}
\end{figure}

The strength and kinematics of the \6\ in this absorber are extraordinary. Integrating over the full profile, this is among the strongest intervening intergalactic \6\ absorbers yet detected, with a rest-frame equivalent width $W_r = 835 \pm 49$ m\AA\ in the $\lambda$1032 profile \citep[cf.][]{Tripp:08:39, Thom:08:22, Danforth:08:194}\footnote{The reported values for some of the other $W_r > 300$ m\AA\ systems vary from study to study, depending on how each system was broken down into components. Our reported value takes in all the identifiable components. This pattern emphasizes the difficulty of interpreting kinematically complex absorbers in terms of simple models, which we will take up below.}. As shown in Figure~\ref{metal_stack}, this absorption spreads over the full 400 \kms\ range of the system, in a nearly flat, complex profile. Figure~\ref{kinplotfig} shows the \6\ profile after it has been converted into apparent column densities using the method of \cite{Savage:91:245}. Accounting for a modest degree of noise in the data, the $\lambda$1032 and $\lambda$1038 profiles agree well over the full velocity range, supporting the case that all this absorption is truly O VI, and that it covers the full kinematic range of absorption by the other detected ionization stages. While component groups A and D appear distinctly in the \6\ profiles, B and C are not clearly separated; they are both blended together and possibly too broad to show individual peaks in the column density plot (Figure~\ref{kinplotfig}). We conclude that \6\ is associated with all four identified component groups in this system.  

Another notable feature of the \6\ is its relationship to the neutral and low-ionization gas in the various components. In component A, strong \6\ at $\log N$(\ion{O}{6}) = 14.6 coincides with strong \ion{C}{3}, \ion{N}{3}, and \ion{Si}{3}, but also \ion{Mg}{2} and \ion{C}{2}. In components B and C, somewhat weaker \6\ is associated with much stronger low ions, such as \ion{C}{2}, \ion{Mg}{2}, \ion{Si}{2}, and \ion{Fe}{2}. In component D, only \ion{H}{1}, \ion{C}{3}, \ion{N}{3}, \ion{Si}{3}, and \6\ are detected, with no corresponding \ion{Mg}{2}. Clearly the state of ionization varies across the absorber, which has an intrinsically multiphase structure.

\subsection{HI Solution}

With the four components defined either by their \ion{H}{1} or metal-line profiles (see notes to Table \ref{comp-list}), we can now derive a solution for the \ion{H}{1} column densities in this absorber. Components A and D both appear in 3 or 4 higher Lyman series profiles, and can be profile-fitted directly. However, both B and C are strongly saturated and blended together, so simple line-profile fitting cannot disentangle them and measure their column densities without additional assumptions. The two straightforward components are taken up first. All relative velocities are given with respect to the adopted systemic redshift of $z_{abs} = 0.3558$. 

{\it Component A:} This component is defined at $-95$ \kms\ by Ly$\eta$, $\theta$, $\kappa$, and $\lambda$, where it appears distinctly but is blended with the blue wing of component B. It appears to be contaminated with unrelated absorption at Ly$\iota$. Profile fits to Ly$\eta$, $\theta$, and $\kappa$ were attempted, with the results appearing in Table~\ref{comp-list}. Consistent results are obtained with $\log N$(\ion{H}{1}) $= 16.42\pm 0.08$, $b =  20 \pm 3$ \kms, at $-94 \pm 1$ \kms. These values are adopted for further analysis. 

{\it Component D:}  This component is defined at $+180$ \kms\ by Ly$\beta$, $\gamma$, and perhaps $\delta$. It is inseparably blended with component C of Ly$\alpha$, contaminated at Ly$\epsilon$, and absent in higher Lyman series lines. Only Ly$\beta$ and Ly$\gamma$ give acceptable fits, with $\log$ N(\ion{H}{1}) $= 14.8$, $b = 16 \pm 3$ \kms\ at 180 \kms. This fit is adopted for further analysis. 

{\it Components B and C:} These two components are both strongly saturated and severely blended with one another at a velocity separation of 60-80 \kms\ (as shown by their metal lines). Because of this blending and saturation we cannot, even in principle, fit them and constrain their column densities separately. Instead, we attempt to impose firm limits on the {\it total} $N$(\ion{H}{1}), based on the Lyman limit opacity and observed Lyman series line profiles.

A firm lower limit to the total $N$(\ion{H}{1}) comes from the observed Lyman-limit opacity. For closely-spaced components, this limit is insensitive to the separations and column density ratios and traces mainly the total column density of \ion{H}{1}. We display two limiting cases in Figure~\ref{wholespecfig}. We adopt the Lyman continuum opacity as function of wavelength given by \cite{1989agna.book.....O}. The red solid curve shows $\log N$(\ion{H}{1}) = 17.8, which leaves $\sim 5$\% transmission across the detected range of the Lyman limit system. By $\log N$(\ion{H}{1}) = 18.0, the transmission is no longer visible and matches the absence of source flux in this region of the spectrum; higher column densities are also consistent with the data. We therefore regard $\log N$(\ion{H}{1}) = 17.8 as a firm, conservative lower limit to the total column density of the LLS. Given the Lyman-limit opacity, 17.8 seems to have a low probability, so we also adopt a slightly higher value of $\log N$(\ion{H}{1}) to bound the lower end of a 95\% confidence interval.

\begin{figure}[!t]
\plotone{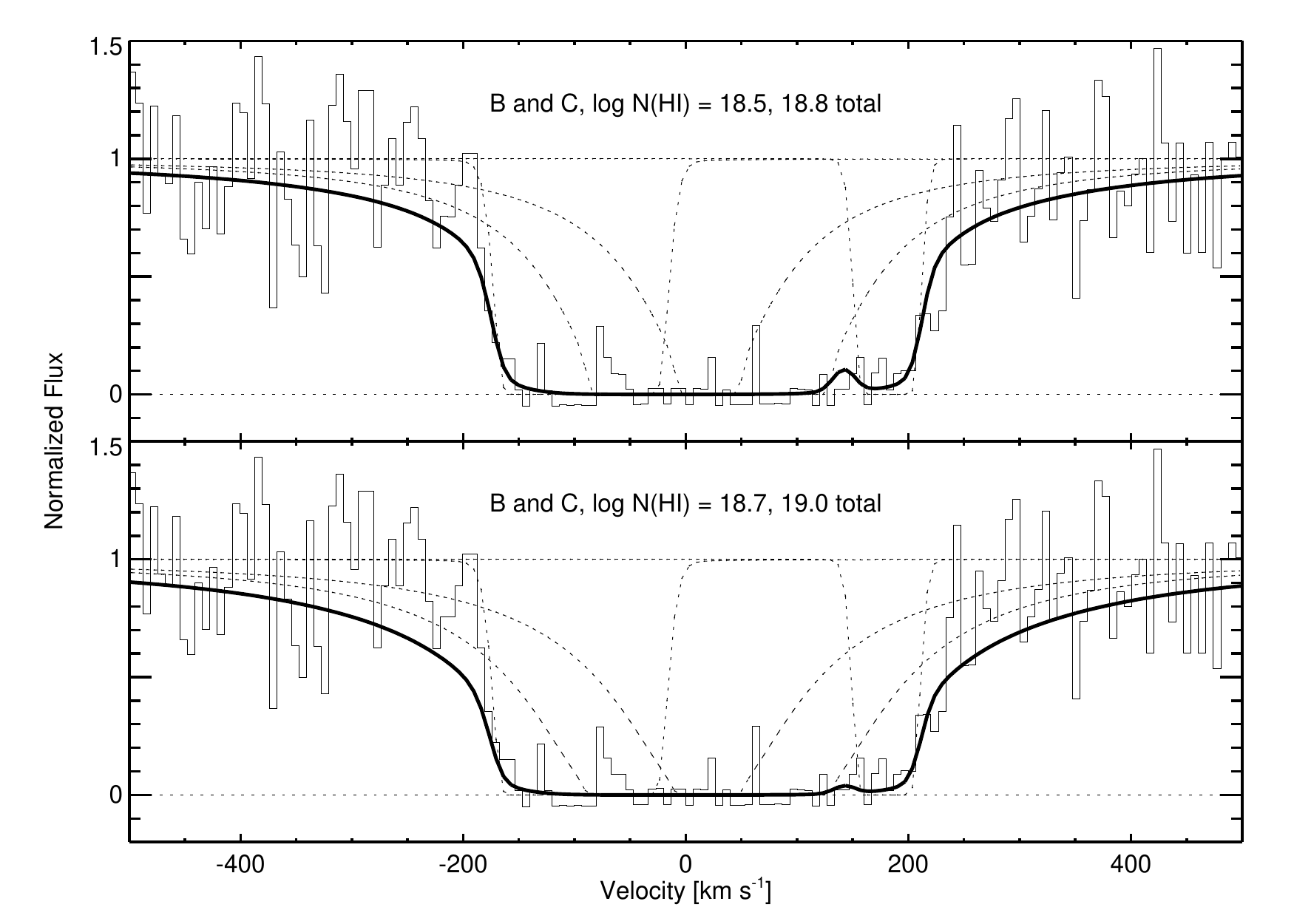}
\caption{The LLS Ly$\alpha$ profile with model lines overlaid, showing the effects of line saturation. The top panel shows a model with $\log N$(\ion{H}{1}) = 18.5 in each of the components B and C, for a total of $\log N$(\ion{H}{1}) = 18.8. The combined profile, shown with the heavy black line, is still a marginally acceptable fit to the data. The lower panel increases the individual components to $\log N$(\ion{H}{1}) = 18.7 and the total to 19.0; by this point damping wings no longer fit the data. \label{lya_stack_fig}}
\end{figure}

Upper limits to total $N$(\ion{H}{1}) come from the absence of damping wings from the Ly$\alpha$ profile of the LLS. This limit is somewhat less firm than the limit from continuum opacity, since it varies slightly with the line broadening and number of components. Two identical components at the positions of the B and C components with $b = 20$ \kms\ and $\log N$(\ion{H}{1}) = 18.5 show damping wings that are only marginally consistent with the data (Figure~\ref{lya_stack_fig}). When each component is increased to $\log N$(\ion{H}{1}) = 18.7 (for a total of 19.0), the damping wings are clearly excluded by the data. These profiles and limits strictly depend on the number of assumed components, which we can infer from the HIRES data. If we assume components at each of the three strongest \ion{Mg}{1} and \ion{Mg}{2} components observed in the HIRES optical data, which are likely to trace the majority of the total \ion{H}{1}, we find that by a total of $\log N$(\ion{H}{1}) $=18.5$, the damping wings appear, and higher values are increasingly poor fits. This is consistent with the two-component analysis. As the number of assumed components within the inner 80 \kms\ increases, the column density per component increases. By $\log N$(\ion{H}{1}) = 19, no tuning of $b$ and number of components (even one) can hide the damping wings, so we regard this value as a firm upper limit to the total N(\ion{H}{1}). The weak presence of damping wings at a total $\log N$(\ion{H}{1}) = 18.8 suggest that the value is likely to be lower than that, so we adopt this as an upper bound to a 95\% confidence interval. For further analysis, we adopt at 95\% confidence interval of $\log N$(\ion{H}{1}) = 18.0 - 18.8, with a flat probability density within that range, since we have no information about which values in this range are more likely.

\section{Metallicity and Star Formation of the Nearby Galaxies}
\label{gal-section}

\begin{figure}[!t]
\epsscale{1}
\plotone{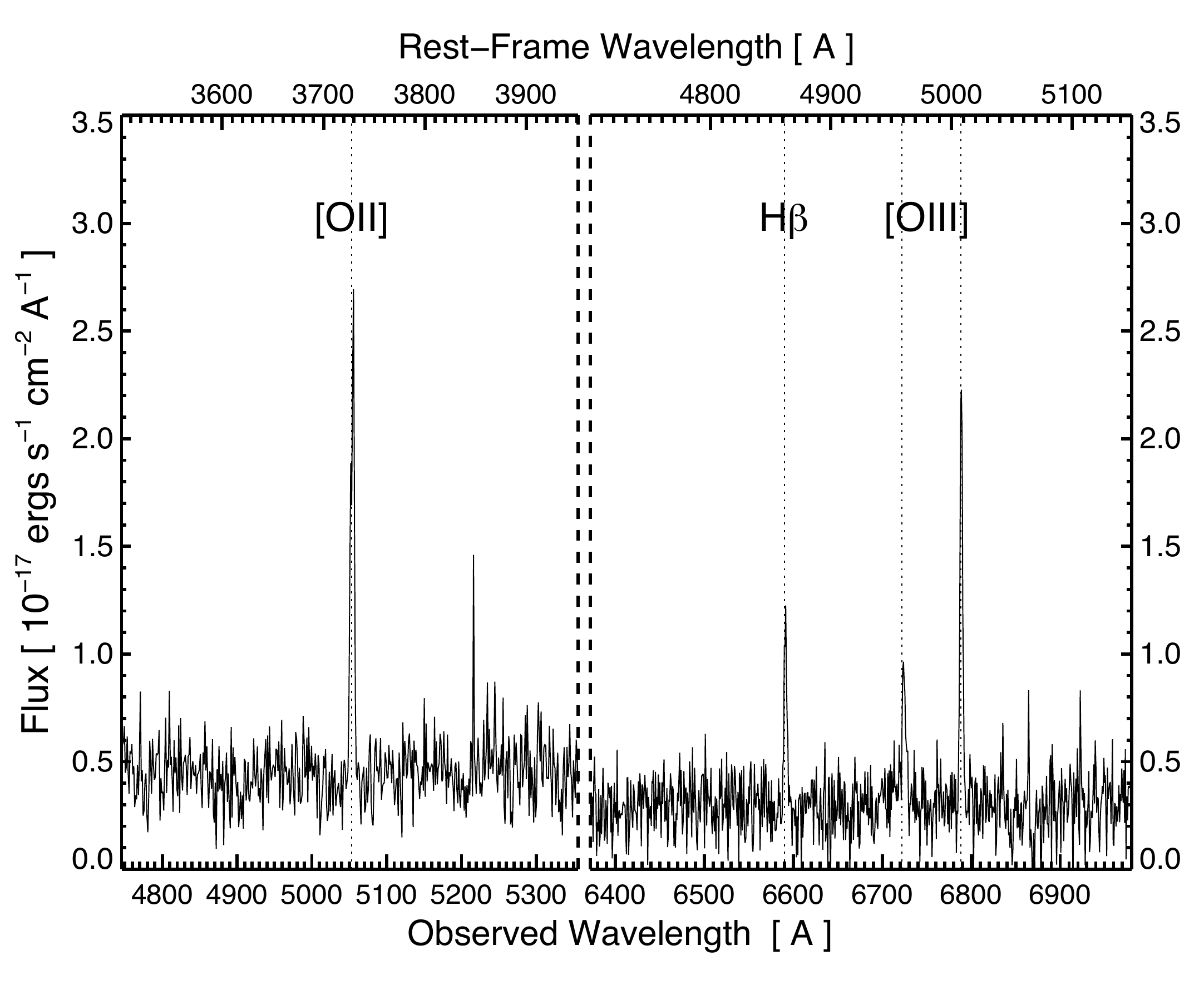}
\plotone{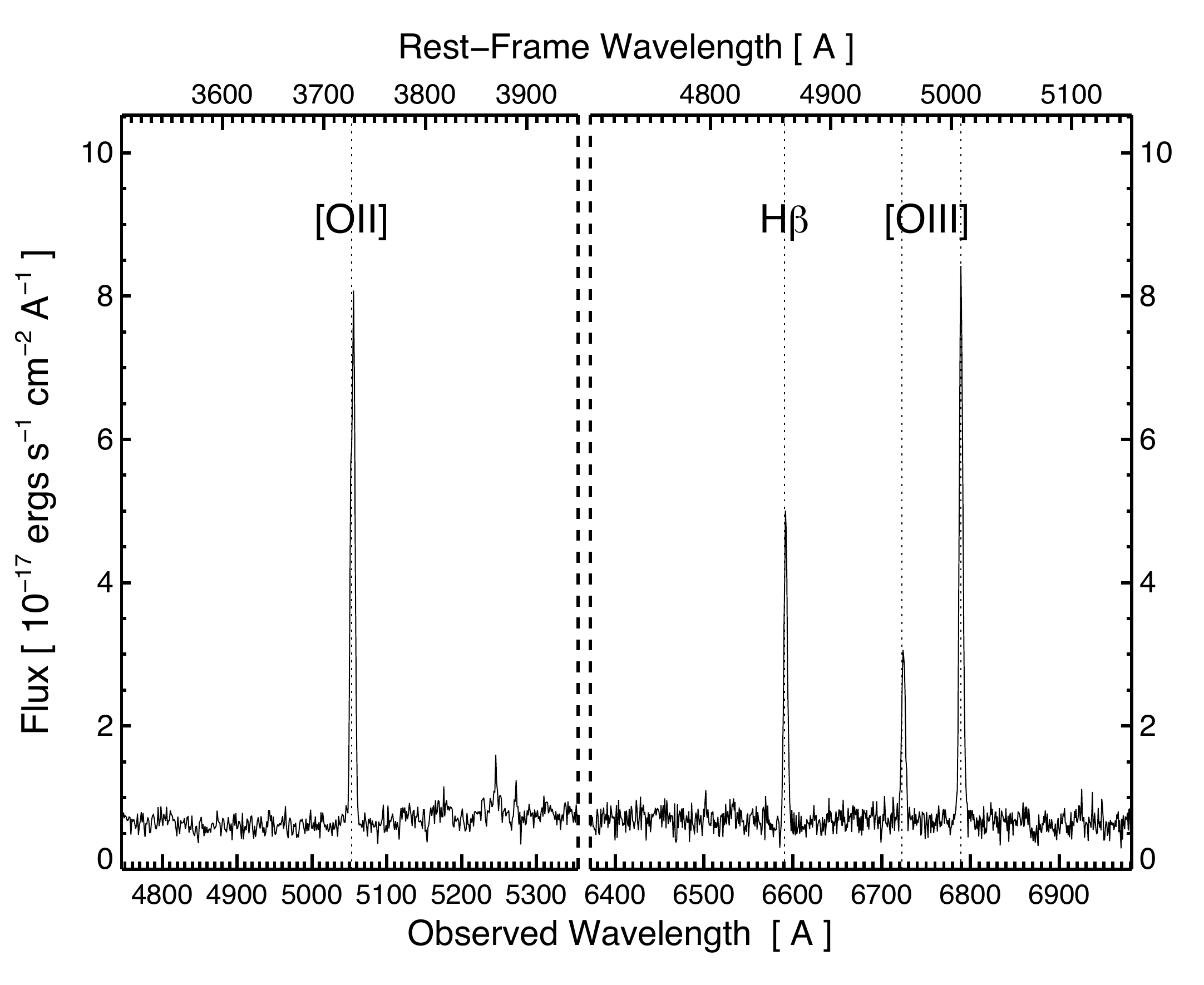}
\caption{Keck/LRIS spectra of galaxies in the field for 86\_4 (top), and 170\_9 (bottom). The break in the wavelength range is placed to avoid LRIS dichroic near 5500 \AA. Detected emission lines are labeled. The red and blue sides for 86\_4 are not normalized as they are for 170\_9 (see \S~2.3). \label{galspec}}
\end{figure}

The Keck/LRIS and WFC3/UVIS data on the galaxies in the field provide useful information about their redshift, luminosity, star formation rate, and metallicity. The reduced Keck/LRIS spectra appear in Figure~\ref{galspec} and the calibrated and PSF-subtracted images appear in Figure~\ref{fieldfig}.

At $z = 0.3558$, [OII] $\lambda\lambda$ 3727, H$\beta$, and [OIII]$\lambda4959$$\lambda5007$ are readily detected. The strengths of these emission lines, with appropriate corrections, provide constraints on the star formation rate and metallicity of the galaxies \citep{Levesque:10:712}. Accordingly, the accuracy of these quantities depends upon the accuracy of the relative emission line strengths and on the degree to which the flux-calibration of the spectra is ``absolute."   We therefore try to ensure that the flux calibrations of the  blue and red sides of the LRIS spectra are consistent with each other, and attempt to correct for light losses in the 1\arcsec\  slit by comparing our spectra with SDSS photometry if it is available for our galaxies. We accomplish this absolute flux correction by convolving the LRIS spectra with SDSS $ugriz$ filters (see \cite{Silva:11:1111} for a description of the IDL code ``spec2mag") and comparing the observed spectral apparent magnitudes with the SDSS catalog apparent magnitudes (see Werk et al., in prep, for details).

Only galaxy 170\_9 has available SDSS photometry that enables us to perform an absolute flux correction.  We do not correct the spectral fluxes of objects 80\_5 and 86\_4 since they do not have SDSS photometry (they are not even detected as one galaxy). The resultant corrective flux factor for 170\_9 is 1.58 (0.494 magnitudes in the $g-$band) on the blue side and 1.65 (0.543 magnitudes in the $i-$band)  on the red side.   Since we have $ugriz$ photometry available to us for object 170\_9, we are able to use the IDL-code $kcorrect$ \citep[][v4\_2]{Blanton:07:734} to obtain an estimate of its stellar mass, approximately 5.5$\times 10^{9}$ M$_{\odot}$.

We estimate the SFR from the H$\beta$ emission line flux using the calibration of \cite{Calzetti:10:1256}, such that SFR (H$\beta$) = SFR (H$\alpha$) /  2.86, where SFR (H$\alpha$) (M$_{\odot}$ yr$^{-1}$) = 5.45 $\times$ 10$^{-42}$ L(H$\alpha$) erg s$^{-1}$. The factor of 2.86 represents the intrinsic ratio of H$\alpha$ to H$\beta$ for case B recombination at an effective temperature of 10,000 K and electron density of 100 cm$^{-3}$ \citep{Hummer:87:801}.  The SFRs appear in Table 2. For 86\_4,  the SFR is a formal lower limit, since we assume there was significant light loss from using a 1\arcsec\ slit (on the order of $\sim$38\%). The errors given in the table account for only the RMS of the line measurements; the flux calibration uncertainty, read noise, sky noise, and flat-fielding errors, are approximately $\sim$1\% for the emission lines in 170\_9 and $\sim$7\% for the lower S/N emission lines in 86\_4.

We obtain the nebular oxygen abundances for galaxies 86\_4 and 170\_9 using the strong line R23 method originally presented by \cite{Pagel:79:95}, according to the calibration of \cite{McGaugh:91:140}.  R23 is defined as log [([OII] $\lambda\lambda3727$ + [OIII] $\lambda4959$ + [OIII] $\lambda5007$)/H$\beta$]. The drawbacks of the R23 method include a well known degeneracy and turnover at $\sim0.3$Z$_{\odot}$ and large systematic errors on the order of 0.25 dex due to age effects and stellar distributions \citep{Ercolano:07:945}.  To correct the emission lines for foreground reddening, we use a reddening function normalized at H$\beta$ from the Galactic reddening law of \cite{Cardelli:89:245} assuming R$_{v}$ = A$_{v}$/E(B$-$V) = 3.1, and using an E(B$-$V) value of 0.013 \citep{Schlegel:98:525}. We cannot break the R23 degeneracy using any of the known methods (e.g. [NII]/[OII]) because our LRIS data do not cover lines redward of $\sim$6500 \AA\ in the rest frame of the galaxies. The upper and lower branch metallicities are presented in Table~2. Based on the well-known mass-metallicity relation of galaxies \citep{Tremonti:04:898}, it is reasonable to assume the upper-branch metallicity for 170\_9 given its estimated stellar mass of $5.5 \times 10^{9}$ M$_{\odot}$. The upper branch value is consistent with solar metallicity according to the updated solar oxygen abundance of  $12 + \log (O/H) = 8.69$ \citep{Prieto:01:L63, Asplund:09:481}.  The spectrum of 86\_4 indicates a slightly lower upper-branch metallicity than the brighter galaxy 170\_9, but given the errors in the measurements, all the upper branch measurements are consistent with solar metallicity and with one another. 

As described below, we unfortunately have no direct measurement of the gas metallicity in the strongest components of the LLS. However, the indirectly indicated metallicity from photoionization modeling is $Z = 0.05 - 0.5 Z_{\odot}$. If this is correct, the gas would appear to have a low metallicity with respect to the galaxies and would favor an explanation in which the gas has entered this system from some other source - the IGM or a stripped dwarf galaxy. However, given the uncertainty in the metallicity of the strong LLS components, and the upper/lower branch ambiguity for the galaxies, we cannot draw firm conclusions from this metallicity comparison. 

\begin{deluxetable}{lccc}[!t]
\tablewidth{0pt}
\tablenum{2} 
\tablecaption{Galaxy Properties} 
\tablehead{
\colhead{Property}&
\colhead{80\_5}&
\colhead{86\_4}&
\colhead{170\_9}}
\startdata
Redshift\tnote{a} 			& 0.88(1) 			& 	0.355574(2)    	& 0.355687(1) 	    \\
$\rho$ [kpc]				& \nodata			&      14.25		& 46.48                  \\
$m_{390W}$\tnote{b}		& $23.1\pm0.2$	& $24.1\pm0.2$	& $22.2\pm0.1$    \\
$m_{625W}$				& $22.5\pm0.1$	& $23.2\pm0.2$	& $21.3\pm0.1$    \\
$[$O/H$]$ (upper)\tnote{c}	& \nodata			& $-0.2$	        	& $0.0$ \\
$[$O/H$]$ (lower)\tnote{c}          & \nodata                   & $-0.6$                      &  $-0.9$   \\
SFR [$M_\odot$ yr$^{-1}$]	& \nodata			& $>0.2$ 	& $2.1 \pm 0.02$
\enddata
\label{galtable}
\tablenotetext{a}{To account for systematic effects in addition to formal fitting errors, we adopt a $\pm 25$ \kms\ uncertainty for these redshifts.}
\tablenotetext{b}{Broadband AB magnitudes derived from the imaging analysis of \cite{Meiring:11:1}.}
\tablenotetext{c}{These estimates are equivalent to [O/H] = $\log$ (O/H) $- \log$ (O/H)$_\odot$, where the solar abundance corresponds to $[$O/H$]$ = 0 or 12 + log(O/H) = 8.69 \citep{Prieto:01:L63,Asplund:09:481}. These estimates carry an error of $\pm 0.15$ dex.}
\end{deluxetable}


\section{Physical Modeling and Interpretation}

In this section we move from measuring column densities and kinematics to interpreting these quantities in terms of physical models, focusing on what we can learn about the ionization, metallicity, and total gas budgets in these components from the detected absorption. 

From the detections of the various ionization stages of C, N, O, Mg, S, Si, and Fe alone it is evident that at least some of the gas in this absorber possesses a high degree of ionization, even in the higher column density component groups B and C. These two components exhibit absorption in the second ions of C, N, S, Si, and Fe, the third ion of S, \ion{O}{6}, and relatively little absorption from the neutrals and first ions of these elements. Apart from any detailed modeling of ionization, the ratios \ion{N}{2} / \ion{N}{1} and \ion{Fe}{3} / \ion{Fe}{2} indicate a significant ionization fraction. The solid detection of \ion{O}{6} in all components further indicates a highly ionized fraction of the gas, whether or or not it arises from the same ionization mechanism as the lower ions. 

To estimate the degree of ionization in the absorbing clouds and from that the total budget of gas that has been detected, we produced photoionzation models with the Cloudy modeling code  (version 08.00) last described by \cite{Ferland:98:761}. This effort is hindered by several limitations that must be acknowledged at the outset. Photoionization models described by a density, ionization parameter, incident spectrum, metallicity, and total gas content require a larger number of uncertain parameters than are formally constrained by the data, so any results must be considered indicative rather than conclusive. The shape and orientation of the absorbing cloud with respect to the nearby galaxies and to the observer are both unknown; we can only assume the simplest cases. While these two factors are generic to absorption systems in QSO sightlines, this particular system presents two additional features that complicate modeling. Its two strongest components, labeled B and C, are inseparably blended with one another in all available lines of \ion{H}{1}, so models are not robustly constrained as usual by a definite N(\ion{H}{1}); we have only a loose constraint on the {\it sum} of the neutral hydrogen column in B and C (see \S~3.5). Furthermore, many of the most important lines of intermediate ions, such as \ion{C}{3} and \ion{Si}{3}, are strongly saturated and blended across all the component groups. 

With these caveats in mind, we model the absorbing clouds with a single plane-parallel slab model illuminated on one side by a uniform incident radiation field. The model field uses two components. First, we attempted models with the composite extragalactic radiation field of Haardt \& Madau (HM05, as implemented within Cloudy). This spectrum yields $\Phi _{HM} = 3.2 \times 10^4$ photons cm s$^{-1}$ for $>1$ Ryd photons at each point in space. By itself, the composite HM field generally fails to achieve the observed ratios of second to first ions of C, N, O, Si, and Fe until the ionization parameter is increased by a low density and the clouds are more than 1 Mpc in size, too large to reside in galaxy halos. It is perhaps not surprising that this spectrum fails, since it represents the average ionizing background of all galaxies and QSOs as transmitted by the IGM. But our absorbers are not located at a random point in space, but within 15 - 50 kpc of galaxies with star formation rates of $\sim 0.2 - 2$ M$_{\odot}$ yr$^{-1}$. Models of the Milky Way halo that attempt to account for the ionization of the Galactic high-velocity clouds typically find that the Milky Way may contribute $10^5$ ionizing photons cm$^{-2}$ s$^{-1}$ even $\sim 100$ kpc into the halo in the direction normal to the disk \citep{Bland-Hawthorn:99:212,Bland-Hawthorn:99:L33, Fox:05:332}, and nearly $10^6$ photons cm$^{-2}$ s$^{-1}$ at 40 kpc\footnote{These models are consistent with the H$\alpha$ emission from the HVCs \citep{Weiner:02:256,Putman:03:948}, but not the Magellanic Stream, which is hypothesized to have at least some of its gas ionized by shocks \citep{Bland-Hawthorn:07:L109}.}. This flux exceeds the contribution of $\Phi _{HM}$ by a significant margin. Thus even a modest degree of star formation, if its UV radiation propagates efficiently into the halo, can exceed the extragalactic background at physical separations like those observed in the present system. 

We therefore add in a second ionizing field component, modeled as a galaxy undergoing a recent episode of star formation, drawn from the Starburst99 libraries \citep{Leitherer:99:3} and corresponding to a continuous star-formation history over 10 Myr with a Salpeter IMF from 1 - 100 $M_{\odot}$ and solar metallicity (the ionizing photon output at $0.5 Z_{\odot}$ is only 15\% higher). For a star formation history extending far into the past, the instantaneous spectrum reaches a steady state beyond a few million years and changes negligibly with the assumed age. The galaxy metallicities (see \S~4) suggest that we should adopt $Z_{\odot}$ for the starburst. This spectrum generates a total of $Q_* = 10^{53.14}$ (SFR / $M_{\odot}$ yr$^{-1}$) ionizing photons s$^{-1}$. We assume that a uniform, wavelength-independent fraction $f_{esc}$ of these ionizing photons escape the ISM of the galaxy to illuminate material in the halo.  At a distance $d$ in kpc, this ionizing field will impart to a cloud of uniform density $n_H$ a field characterized by the dimensionless ionization parameter $U$:  
\begin{align*}
  U &=  \frac{1}{c n_H}  \left( \frac{Q _* f_{esc}}{4 \pi d^2}  + \Phi _{HM} \right) \\
       &=   \frac{0.01\, {\rm cm}^{-3}}{n_H} \left[ 10^{-3}  \frac{SFR}{M_{\odot} {\rm yr}^{-1}}  \frac{f_{esc}}{0.1}  \left( \frac{20\, {\rm kpc}}{d} \right) ^2 + 10^{-4} \right]
\end{align*}
where in the final equation the quantities have been scaled to typical values for halo clouds. This relation does not account for radiative transfer in the Lyman continuum, and assumes normally incident radiation, and so it is only an approximation for real clouds. For $d = 20$ kpc and $SFR = 1$ M$_{\odot}$ yr$^{-1}$, the ionization parameter $\log U \sim -3$. The constants $10^{-3}$ and $10^{-4}$ are convenient approximations good to better than $10$\%. In these typical conditions, the stellar ionizing flux exceeds the HM spectrum by a order of magnitude, consistent with results from the Milky Way. At all higher SFRs, higher $f_{esc}$, and lower distances, the stellar ionizing field is even more dominant. Since we do not know the cloud distance to the galaxy, the density, or the $f_{esc}$, we cannot use this relation to specify $U$ for a given absorber. However, once a photoionization model has been found and $U$ constrained by observations, along with SFR for the galaxies, we can use this relation to obtain estimates of the other parameters. 

The sum of these HM05 and starburst radiation field components is normalized within Cloudy to achieve a specific ionization parameter at the face of the model cloud. The model clouds are calculated from their sharp illuminated face to a stopping column density of $\log N$(\ion{H}{1}) = 19.5, to cover components B and C; these models are valid for any lower total $N$(\ion{H}{1}) if their results are truncated at that column density so are applicable to A and D as well. The model clouds are assumed to have a uniform metallicity of appropriately scaled solar relative abundances; models with total metallicity from solar to 1\% solar were explored. 

\subsection{Components B and C}

\subsubsection{Metallicity}

In B and C we detect weak absorption from neutral oxygen in two transitions: \ion{O}{1} $\lambda$1302 and $\lambda$988. These lines are likely unsaturated, as the apparent column density profiles shown in Figure~\ref{kinplotfig} (middle panel)  are consistent in the strong component groups B and C. The B component of the $\lambda$1302 line appears mildly contaminated; we adopt the value $\log N$(\ion{O}{1})$ = 14.5 \pm 0.2$, boosting the error from 0.1 to 0.2 dex to account for the uncertainty in the contamination. For component C, we obtain consistent measurements from both lines, $\log N$(\ion{O}{1}) $= 14.2 \pm 0.2$. Significant absorption is not detected in components A or D, and direct integrations over their velocity ranges yield upper limits of $\log N$(OI) $< 13.9$ ($2\sigma$). 

Neutral oxygen is a useful metallicity indicator in moderately ionized gas, since its ionization potential of 13.6 eV and charge exchange reactions with H lock its ionization fraction to that of hydrogen in typical conditions. Unfortunately these well-detected \ion{O}{1} lines must be compared to a highly uncertain total \ion{H}{1} column, ranging anywhere from $\log N$(\ion{H}{1}) = 18.0 - 18.8 (95\% confidence).  With this range, metallicities from $0.1 - 1.0 Z_{\odot}$ are all permitted in comparison with the updated solar oxygen abundance $12 + \log (O/H) = 8.69$ \citep{Prieto:01:L63, Asplund:09:481}. We therefore cannot obtain a robust, direct measurement of gas metallicity in this system, owing to the systematic error in $N$(\ion{H}{1}) introduced by the strong saturation and blending of the two inner components. However, we can impose upper and lower limits, $Z = 0.15 - 1.0 Z_{\odot}$ based on the 95\% confidence range derived for $N$(\ion{H}{1}) above (or $0.1 - 1.6 Z_{\odot}$ with the firm limits on $N$(\ion{H}{1})). This circumstance somewhat limits our ability to interpret this material as either infall or outflow from the associated galaxies. However, modeling of absorption from ionized gas indirectly indicates sub-solar metallicity, as shown in the next section.

\subsubsection{Ionization}

\begin{figure}
\epsscale{1}
\plotone{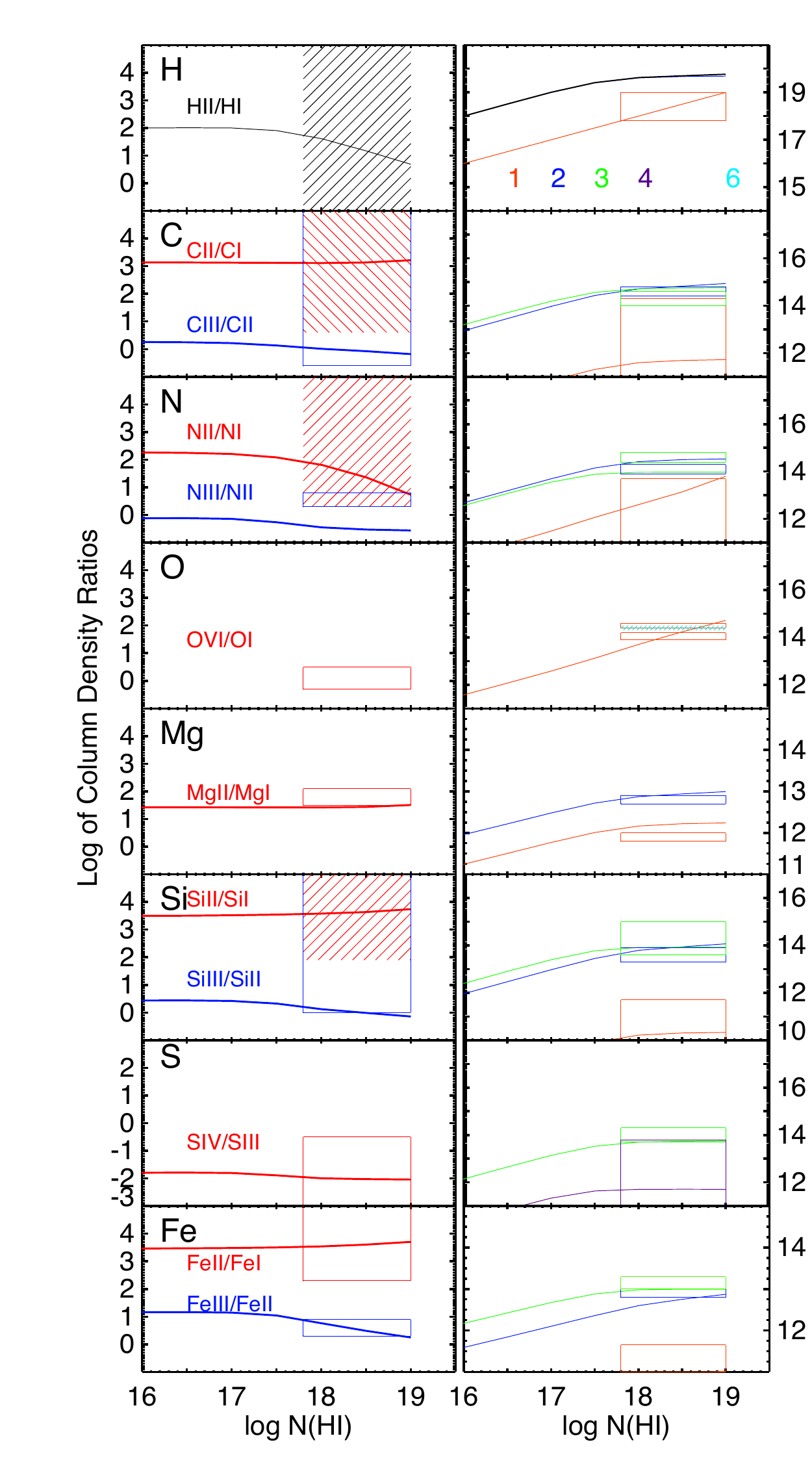}
\caption{Photoionization modeling for the B and C velocity components of the LLS. This model assumes the Starburst99 spectrum described in the text with $\log U = -3.5$ and $Z = 0.1 Z_\odot$. The left column shows column density ratios of key ions, while the right shows the cumulative column densities. Both are plotted with respect to cumulative N(\ion{H}{1}). The ranges permitted by the data for column densities and ratios are marked, coded in color to match their model curve. In the right column, the species are color coded according to their stage of ionization, e.g. 1 equals \ion{C}{1} in orange, 2 equals \ion{C}{2} in blue. \label{allions}}
\end{figure}

The results of our Cloudy photoionization model with this Starburst99 input spectrum appear in Figure~\ref{allions}, specifically for the parameters $\log U = -3.5$, and 10\% solar metallicity in the gas. The left column of this figure shows the column density ratios of the key ions, and the right column shows the cumulative column densities, $N$(X), as a function of \ion{H}{1} column density into the cloud (note that this is not physical depth, or total H column, which is proportional to depth for constant density). These quantities are plotted with respect to a varying $N$(\ion{H}{1}) since we must account for the uncertain total \ion{H}{1} column density. If that column density were well-constrained by measurement, we could choose one vertical slice though this space with a fixed $N$(\ion{H}{1}) and analyze models of varying $U$ and $Z$, as is typically done for photoionization models. Instead, we analyze these families of models with a range of $N$(\ion{H}{1}) and proceed in two steps. First, we adjust the ionization parameter $U$ to place the ion ratios in the left column within their permitted ranges. Then, holding $U$ fixed, we vary the metallicity until the cumulative column densities in the right column fall into the observed ranges for themselves individually and also for \ion{H}{1}. Where these curves lie inside the boxes that mark data, both the total $N$(X) and the $N$(\ion{H}{1}) are reproduced by that particular model. Thus, the ion ratios constrain the ionization parameter while the column densities constrain the metallicity. 

The model with $\log U = -3.5$ provides an acceptable fit to the ratios of the observed ions, except that it underproduces \ion{N}{3} and produces essentially no \ion{O}{6} or \ion{S}{4}. However, the detected \ion{N}{3} is likely contaminated by \ion{Si}{2} or other absorption and so the observed ratio \ion{N}{3} / \ion{N}{2} is artificially high; we discount this ratio to discriminate models. Otherwise, these parameters are a good model for the components B and C. At the depth into the cloud at which this model has achieved $N$(\ion{H}{1}) $\simeq 18 - 18.8$ the clouds are still predominantly ionized. As shown in Figure 8, the lower $N$(\ion{H}{1}) models within the permitted range have lower neutral fractions, while the higher $N$(\ion{H}{1}) have higher neutral fractions, so that the typical total N(H) range of these models is $N$(H) $\simeq 19.5 - 20$. Thus in models with pure photoionization, these clouds are nearly all ionized. 

The column densities of well-measured ionization stages (right column) also indicate a subsolar metallicity. This constraint on the metallicity should be considered indirect, because it depends on the assumption of pure photoionization. Yet under this assumption, models with an ionization parameter, $\log U \simeq -3.5$,  that matches the observed ratios of ions from the same element fail to match the column densities of all these ionization stages unless the metallicity is $5 - 50$\%, as shown in Figure~\ref{allions}. This indirect measure of the gas metallicity is affected less by the saturation of \ion{H}{1} than might be expected. As shown in the right column of Figure~\ref{allions}, the column densities of the first and second ions increase slowly with respect to $N$(\ion{H}{1}) through the range where it increases from 18.0 to 18.8, because over this range the H is becoming increasingly neutral and is added to the line integral through the cloud at a much higher rate, per pathlength, than species such as \ion{Mg}{2} or \ion{Fe}{2} for which the ionization fractions change little over this range. For a given $U$, these cumulative column density curves shift up and down in those panels of the figure in direct proportion to the assumed cloud metallicity, which has been adjusted to best match the greatest number of column densities. The best fitting model has $0.1 Z_{\odot}$, but values from $0.05 - 0.5 Z_{\odot}$ still provide decent matches for most ions. Even though the metallicities estimated directly from the detection of \ion{O}{1} are highly uncertain because of the saturation of the \ion{H}{1}, they are consistent with these models at the higher end of the permitted range of $N$(\ion{H}{1}). 

Finally, we have assumed a nominal density of $n_H = 0.01$ cm$^{-3}$, which yields cloud sizes of $1 - 3$ kpc. We therefore have a model in which the two strongest components in the LLS absorber are consistent with diffuse, possibly low-metallicity clouds of total gas column $\log N \simeq 19-20$ and with kpc scale that are ionized by the stars in their nearby galaxies. These absorbers suggest that star-forming galaxies with $f_{esc} = 0.1$ may be able to photoionize some of their own halo gas under the right conditions. 

In light of the appearance of many ionization stages in this absorber, e.g. both \ion{O}{1} and \ion{O}{6}, we must consider that some or all of the gas arises in warm or hot collisionally ionized material, such as might arise in a shock or conductive interface between hot and cold gas. For purely collisionally ionized gas without a photoionizing field, we can use individual ionization ratios to constrain the temperature. The ratios of the second ions to the first ions of C, N, Si, and Fe  (e.g. \ion{N}{3} to \ion{N}{2}) all prefer a temperature of between $20000 - 80000$ K. Over this range the neutral fraction of H drops from $\sim 0.1$ to 10$^{-4}$, so if collisional ionization dominates in these absorbers, the total column density of H will exceed $\log N$(\ion{H}{1}) = 19.5 and could be substantially higher. However, the observed ratio of \ion{Mg}{2} /  \ion{Mg}{1}  = 1.8 dex prefers a temperature $T < 20000$ K, and is one of the better determined ratios in components B and C. Thus, no single temperature in CIE describes all the detections. By contrast, the photoionization model above appears to be able to account for these ratios altogether, with the significant exception of the \ion{O}{6}. 

An important caveat to this modeling is that we have considered ionization mechanisms in isolation, rather than possible combinations of them. Such a model might explain the co-existence of \ion{Mg}{2} and \ion{O}{6}, for instance, if the latter is collisionally produced while the former arises in a photoionized cloud. Such a model will require more parameters than either model on its own, but may be tenable. We will discuss the possible ``multiphase'' nature of the \ion{O}{6} absorber below. 

Indeed it is impossible to reconcile \ion{O}{6} and low ions such as \ion{Mg}{1} into the same CIE or photoionization model, and both of these appear in components B and C. Since the \ion{O}{6} is not readily reproduced by either of the photoionization models considered above, it probably does arise in collisionally ionized gas. If so, the corresponding \ion{H}{1} may be too weak to separate from the stronger \ion{H}{1} from the photoionized component. If the \ion{O}{6} exists at its peak ionization fraction $f_{OVI} = 0.2$ in CIE at $T = 10^{5.5}$ K, the corresponding \ion{H}{1} has N(\ion{H}{1}) $\simeq (2.5/Z) \times 10^{12}$ cm$^{-2}$ where $Z$ is the metallicity relative to solar (thus $2.5 \times 10^{13}$ cm$^{-2}$ at 0.1$Z_{\odot}$). Only if the assumed temperature drops to $100,000$ K, where $f_{OVI} \simeq 10^{-5}$, does the associated \ion{H}{1} exceed $10^{18}$ cm$^{-2}$ and so become comparable to the observed values. However, at this temperature the H neutral fraction is also $\simeq 10^{-5}$ and the total H column is an implausibly large $10^{23}$ cm$^{-2}$. We therefore conclude that a collisional origin for the \ion{O}{6} (and possibly other high ions for which we do not have coverage) is possible, and that the corresponding \ion{H}{1} would go undetected under most circumstances. Such a case might occur if the \ion{O}{6} arises in the interface layer between a photoionized cloud and a hot, diffuse halo, as has been proposed for the high-velocity cloud population of the Milky Way \citep{Sembach:03:165, Fox:04:738}. As these interfaces lie at intermediate temperatures between hot and cold material, they may account for some portion of the intermediate ionization stages as well. Were we able to account for this properly, it might change our conclusion about the photoionization models described above. However, these interface models typically produce less \ion{O}{6} than is observed, as discussed more below.  

We are therefore forced to conclude that these components consist of complex multiphase material, and that they may arise from some combination of photo- and collisional-ionization. The essential point of these ionization modeling exercises is not that a particular model gives a perfect fit to the observed ion ratios; the uncertain N(\ion{H}{1}), component blending, and line saturation ensure that no perfectly robust modeling is tenable. However, ionization models with a range of plausible conditions all point in the same direction; that the bulk of the gas traced by the \ion{H}{1} and the metal-line absorption is highly ionized, with a total H column density exceeding that seen in \ion{H}{1} by a factor of 10 - 100. There is still more highly ionized gas, traced by \ion{O}{6} in all the components, that may indicate interfaces between this photoionized halo gas and its environment. While we cannot completely rule out collisional ionization for the bulk of the strong components B and C, a photoionization model incorporating UV light from star forming regions in the nearby galaxies appears plausible if the cloud has density $n_H \simeq 10^{-2}$ cm$^{-3}$, lies 20 - 50 kpc from the associated galaxies,  and if those galaxies propagate $\sim 10$\% of their ionizing photons into the halo. 

\subsection{Component A}

\begin{figure}[!t]
\epsscale{1.2}
\plotone{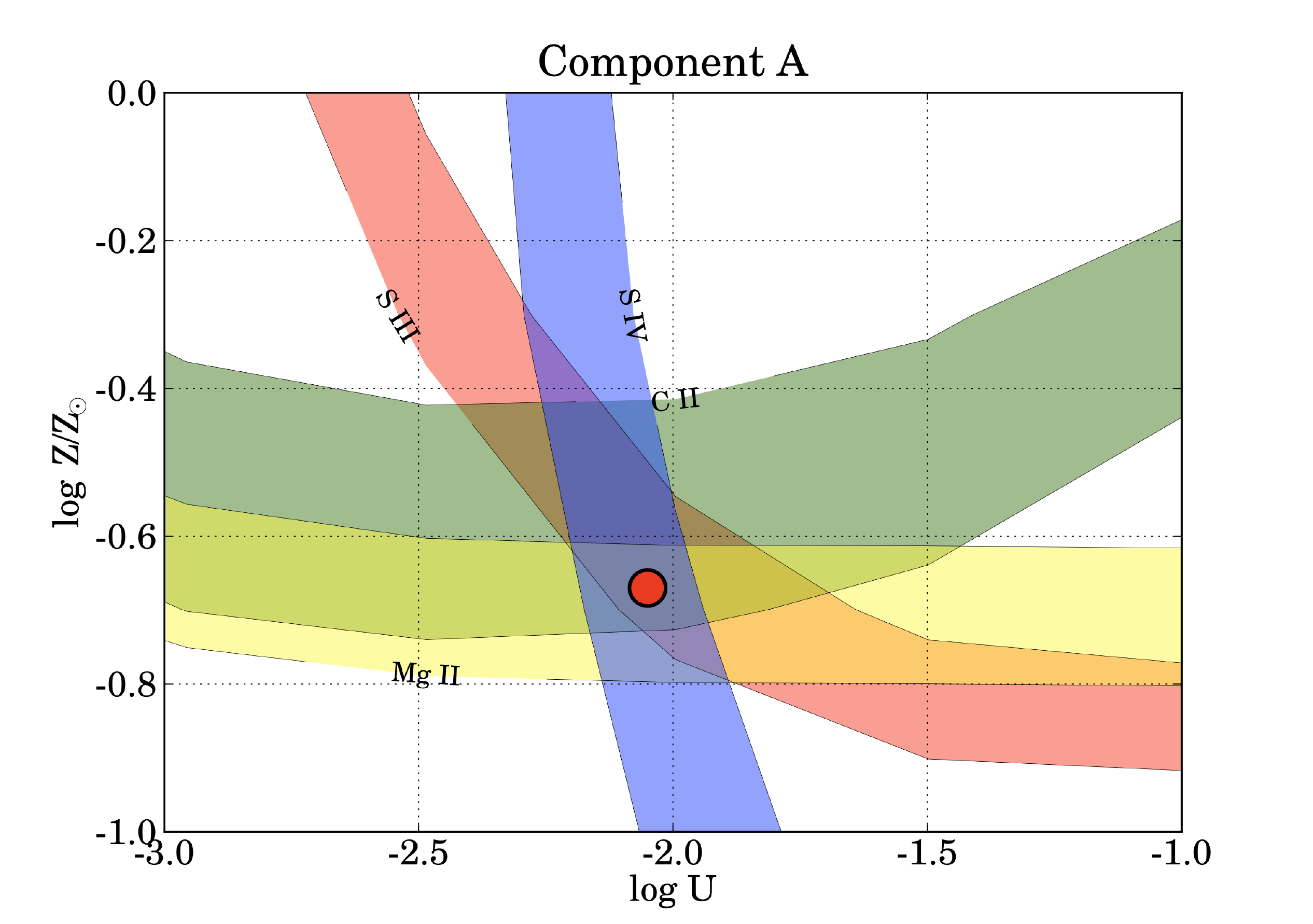}
\caption{Parameter space for photoionization models corresponding to Component A. The most informative constraints are imposed by the column densities of \ion{C}{2} (blue), \ion{Mg}{2} (orange), \ion{S}{3} (black), and \ion{S}{4} (red).   \label{compAfig}}
\end{figure}

This component exhibits a wide range of ionization stages, from \ion{Mg}{2} to \ion{O}{6}. This range of species cannot be accommodated within a single photoionization or collisional ionization model, for the same reasons that applied to B and C above; it is clearly ``multiphase''. Following the same procedure as above, we attempted to fit as many species as possible into a single model to estimate what properties the cloud might have under various physical scenarios. 

We first ask whether the combination of a galactic star-forming spectrum with the diffuse extragalactic background will provide a suitable model, as it did for components B and C. Since we have a robust measurement for $\log N$(\ion{H}{1}) = 16.5 in this case, we can restrict our attention to models that give that value. These results appear in Figure~\ref{compAfig}. Most regions of this parameter space with $\log U \gtrsim -3$ (from the combination of the HM and S99 spectra) satisfy the lower limits on \ion{C}{3} and \ion{Si}{3}. The measured column densities of \ion{C}{2},  \ion{Mg}{2},  \ion{S}{3}, and  \ion{S}{4} are constraining, and mutually consistent in a small region centered around the filled circle at $\log U = -2$, $Z = 0.2$ $Z_{\odot}$. This model has a total H column of $\log N$(H) $\simeq 19$, and thus a neutral fraction of only $10^{-2.5}$. However, the ionization parameter is 10 - 30 times higher than for B and C, suggesting that this component has some combination of lower density and/or proximity to the ionizing source. We therefore conclude that this photoionizing model is plausible for all the species except the \ion{O}{6}. 

The \ion{O}{6} detection suggests that some of this gas may reside in a hotter, collisionally ionized component. We pursued this hypothesis by adjusting the temperature and metallicity of a purely collisionally ionized model cloud to achieve matches to the observed ratios and column densities. No single temperature is clearly indicated; the well-measured ratio of \ion{S}{4} to \ion{S}{3} is matched by a model with 80000 K, but this model greatly underproduces both \ion{Mg}{2} and \ion{O}{6}, which has its highest ionization fraction at $\sim 300000$ K. Single-temperature models that give \ion{O}{6} column densities above $10^{14}$ cm$^{-2}$ typically give \ion{Mg}{2} column densities 100$\times$ lower than we measure here. A model with $\log T = 5.3$ gives a ratio of \ion{O}{6} to \ion{S}{4} that is near the observed value, but overpredicts the observed ratio of \ion{S}{4} to \ion{S}{3}. Where therefore conclude that no single-temperature collisional ionization model can explain the absorption by component A; distinct models are needed for the low ionization gas up to \ion{C}{3}, \ion{Si}{3}, and \ion{S}{3}, while a hotter, collisionally ionized phase could account for the \ion{O}{6} and possibly the \ion{S}{4}. In a combined model, the neutral H fraction associated with the collisional component is likely quite low with respect to the measured $\log N$(\ion{H}{1}) = 16.5, which would trace predominantly the photoionized component.  We thus have no constraint on the metallicity of the hotter component.

\begin{figure}[!t]
\epsscale{1.2}
\plotone{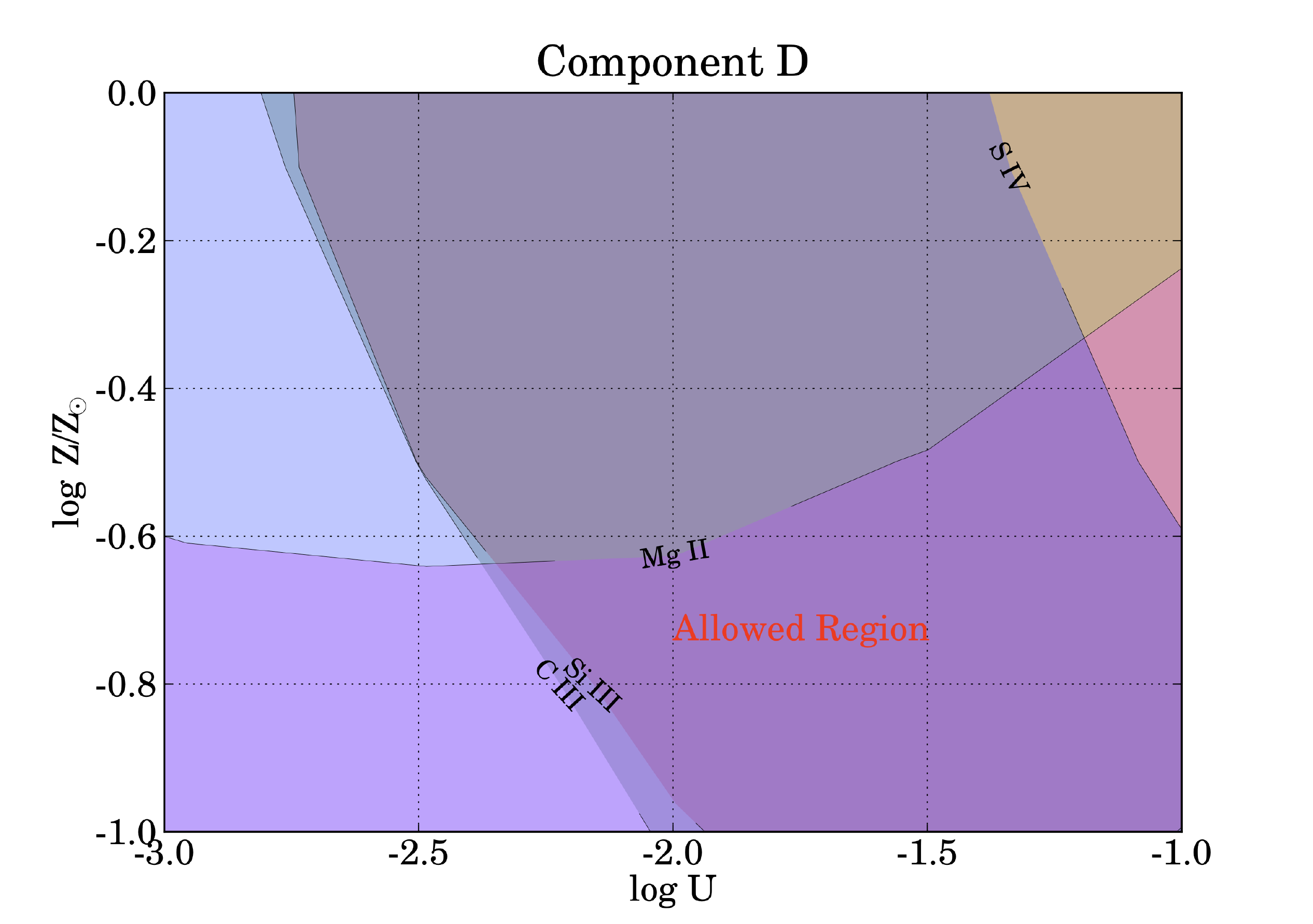}
\caption{Parameter space for photoionization models corresponding to Component D. The strong detections of \ion{C}{3} and \ion{Si}{3} exclude models to the left of the green and blue lines, while the upper limit on \ion{Mg}{2} requires models below the orange line and the upper limit on \ion{S}{4} requires models to the left of the red line.  \label{compDfig}}
\end{figure}

Thus we find that Component A can also be accommodated into a simple scenario in which low-metallicity gas in the halo(s) of the nearby galaxies is ionized by their star formation. Such a simple scenario can account for the observed column densities with the significant exception of the \ion{O}{6}, for which a collisionally ionized phase may be preferred. 

\subsection{Component D}

Component D has a column density $\log N$(\ion{H}{1}) = 14.8, and exhibits strong absorption in \ion{O}{6}, saturated \ion{C}{3} and \ion{Si}{3}, and the problematic transition of \ion{N}{3}. Significantly, no \ion{Fe}{3} is seen here as in B and C. With fewer detected ions, models are more loosely constrained. In this case, a single temperature CIE model can account for the column densities of \ion{C}{3}, \ion{Si}{3}, \ion{N}{3}, and \ion{O}{6} in a narrow temperature range around $\log T \simeq 5.25$, provided the total H column density is $\log N$(H) $\sim 20$ and the metallicity is solar. However, this model conflicts with the observed doppler parameter $b = 16$ \kms\ obtained by consistent fits to the profiles of Ly$\beta$ and Ly$\gamma$. At $\log T = 5.3$, the doppler parameter for H should be no less than $b = \sqrt{2kT/m_H} = 57$ \kms, but the line profiles prefer a value $T \lesssim 20000$ K, which corresponds to $b = 16$ \kms. We therefore rule out a single CIE model for all the detected absorption in this component. 

If we consider a photoionization model for this component, we find that the number of detected ions is generally insufficient to constrain the ionization parameter to the degree that was possible for A, B, and C. Any model with $\log U \gtrsim -2.5$ gives the correct column densities of \ion{C}{3} and \ion{Si}{3}, but tends to produce too much \ion{Mg}{2}, which is undetected, unless the metallicity is less than $0.3$ solar. This region of parameter space is shown in Figure~\ref{compDfig}. The allowed area with $\log U = 2.5$ to $-1$ and below $\sim 25 - 40$\% solar produces enough \ion{C}{4} and \ion{Si}{3} without producing too much \ion{Mg}{2} or \ion{S}{4}. The total gas column densities N(H) range from $10^{18} - 10^{19.5}$ cm$^{-2}$ as $\log U$ increases from $-2.5$ to $-1$. This component would appear to have a very large ionization correction, with only $10^{-3}$ or $10^{-4}$ of the hydrogen in \ion{H}{1}. As for the other components analyzed above, we find that the combined ionizing radiation can explain the detected absorption, but only if the gas is highly ionized and has a subsolar metallicity. 

\subsection{What Creates the \ion{O}{6}?}

\label{o6section}

None of the analysis just presented answers convincingly the question of what creates the highly ionized gas traced by the \ion{O}{6}. Models with purely photoionized gas do not produce enough \ion{O}{6}, or if they do they do not also explain the lower-ionization gas. It is the coexistence of this strong \ion{O}{6}, together with the low ions, that make this absorber ``multiphase''. 

We must consider the possibility that the highly ionized gas in this system -- the \ion{O}{6} and perhaps some or all of the \ion{S}{4}, \ion{S}{3}, and \ion{N}{3} -- arise in non-equilibrium situations such as shocks, conductive interfaces between hot and cold material, or simply cooling gas which is out of thermal and/or ionization equilibrium. \cite{Fox:04:738} and \cite{Indebetouw:04:205, Indebetouw:04:309} have compiled models for many of these phenomena, and mapped out their predicted column-density ratios in \ion{Si}{4}, \ion{C}{4}, \ion{N}{5}, and \ion{O}{6}. Unfortunately our COS spectra have coverage of only the latter two ions, and \ion{N}{5} is not detected. The ratio N(\ion{N}{5}) / N(\ion{O}{6}) provides only an upper limit for the four component groups, which ranges from $< -0.6$ to $< -1.0$. These limits provide no constraint, since they are consistent with shock ionization, radiative cooling, conductive interfaces, and turbulent mixing layers, as compiled by \cite{Fox:04:738}. The uncertainty of comparing models to data in this fashion is complicated by the possibility that the gas under study has non-solar relative abundances of the diagnostic elements C, Si, N, and O, which allows the regions occupied by these various models to move within the column density ratio space. While \cite{Fox:04:738} are able to correct these abundance ratios based on independent measurements of relative abundances, we do not have that much constraint here. So we are unable to produce quantitative constraints on ionization mechanisms using this technique. 

This absorber is notable not only for its very high total column density on \ion{O}{6}, but also for how it produces this high value. We do not see a monolithic, saturated component in \ion{O}{6}, thermally broadened into a single profile in hot gas, nor do we see \ion{O}{6} associated only with the strongest or weakest \ion{H}{1}. Rather, we see a complex profile in which the \ion{O}{6} absorption is spread over nearly 400 \kms\ and breaks into at least two and possibly more distinct components, which line up with components at similar velocities in lower ions. With a sightline that extends over $> 100$ kpc through the halos of two galaxies, it seems natural to guess that the \ion{O}{6} arises in multiple distinct objects within the halo which may or may not be associated with the individual galaxies. Strong shocks ($v \sim 600-2000$ \kms) can exhibit column densities in the range of our detections \citep{Gnat:09:1514}, but imply post-shock temperatures of $T \sim 5 \times 10^6$ K and \ion{H}{1} thermal linewidths $b \gtrsim 70$ \kms. The strength and complex component structure of the \ion{H}{1} lines precludes a search for such a broad component, which would only be detectable in the Ly$\alpha$ profile in this Lyman-limit system.  Thus while we cannot rule out shocks for some of this \ion{O}{6}, we have no strong indication for them either.  We note that some \ion{O}{6} absorbers reported in the literature have corresponding broad \ion{H}{1} components that could indicate the expected temperature in shocked gas \citep{Tripp:01:724, Richter:04:165, Narayanan:10:1443, 2011arXiv1102.2850S}.

One possible explanation for the observed kinematic and ionization pattern is that the \ion{O}{6} arises in transition layers between hot and cold gas, such as might occur for neutral or photoionized clumps falling through a Galactic halo and interacting with the hot halo gas left there by the formation of the galaxy.  \ion{O}{6} is believed to trace interfaces between stable cold ($T \sim 10^4$ K) and hot ($T > 10^6$ K) phases of the ISM and circumgalactic medium because it achieves its maximum ionization fraction ($T = $100-300,000 K) near the peak of the radiative cooling curve and so is short-lived. In this simple scenario, interfaces arise where cold clouds interact with the hot medium and are generally unrelated to, or weakly dependent upon, the size and/or quantity of gas in the cold clumps. Most simply there is just one interface per cloud, however large the cloud. 

Quantitative models of conductive interfaces and turbulent mixing layers typically produce column den fsities of \ion{O}{6} in the range $10^{12-13}$ cm$^{-2}$, lower than we detect here \citep[cf.][]{Borkowski:90:501, Slavin:93:83, Indebetouw:04:205, Indebetouw:04:309, Gnat:10:1315}, and so multiple interfaces must be invoked. The models of \cite{Gnat:10:1315} find that \ion{O}{6} column densities typically have N(\ion{O}{6}) $< 10^{13}$ cm$^{-2}$ for cold clouds bounded by conductive interfaces within a hot corona, so that $\sim 100$ interfaces are implied by the total column density of the J1009+0713 absorption-line system. A very large number of low-column density \ion{O}{6} components probably could be accommodated by our data, since blending and thermal broadening would likely make them difficult to distinguish from a smaller number of higher column density components, which is which is similar to what we observe. 

However, the interface scenario implies a simple observational pattern: that \ion{O}{6} should appear whenever low ions trace colder clumps in the halo, and that the strength of \ion{O}{6} absorption near galaxies should increase with the number of detected components in the cooler material, and correlate with them in velocity space. The overall pattern observed so far is that is that the column density of \ion{O}{6} is relatively insensitive to the total $N$(\ion{H}{1}) in surveys of IGM absorbers,  though the patterns depend sensitively on the number of defined components, which vary from study to study especially for the strongest absorbers.  The column density of \ion{O}{6} varies in our case by only about 0.2 dex across the four detected component groups, while the \ion{H}{1} column density varies by at least three and possibly four orders of magnitude. This striking lack of correlation suggests that the presence and total quantity of \ion{O}{6} is not related directly to the total quantity of gas. Indeed, this is a generic pattern of \ion{O}{6} absorbers that has been noticed in systematic surveys \citep{Danforth:08:194, Tripp:08:39, Thom:08:22}. In the compilation of \ion{O}{6} absorbers by \cite{Thom:08:22}, $N$(\ion{O}{6}) varies by 1.5 dex over 5 decades of variation in $N$(\ion{H}{1}). This is partly due to the dramatically multiphase character of some of the absorbers in the Thom \& Chen sample, but even the ``simple'' \ion{O}{6} absorbers (systems with well-aligned \ion{O}{6} and \ion{H}{1} components and simple component structure with no clear evidence of multiple phases) identified by \cite{Tripp:08:39} show this effect, with N(\ion{H}{1}) varying by 3 orders of magnitude while N(\ion{O}{6}) only changes by 1 dex. Taken as a system, the J1009+0713 LLS occupies the extreme upper right of this diagram, with $\log N$(\ion{O}{6}) = 15.0 at $\log N$(\ion{H}{1}) $\sim 18.5$, but taken by components the four groupings scatter over the full range found by \cite{Thom:08:22}. Thus the comparison of \ion{O}{6} to \ion{H}{1} is roughly consistent with the interface scenario.

\begin{deluxetable*}{ccccccc}[!t]
\tablenum{3} 
\tablecaption{Literature on OVI-bearing LLSs} 
\tablehead{
\colhead{\#}&
\colhead{Sightline}&
\colhead{$z_{abs}$}&
\colhead{$\log N$(\ion{H}{1})}&
\colhead{$\log N$(\ion{O}{6})}& 
\colhead{$N_{comp}$}& 
\colhead{Reference} }
\startdata
1 & HE0153-4520	& 0.022601 	&	$16.61^{+0.12}_{-0.17}$  & $14.21 \pm 0.01$			& 1 & 1   \\
2 & PKS0312-77	& 0.2026		&     	$18.22^{+0.19}_{-0.25}$	& $14.95 \pm 0.05$			& 6 & 2 \\
3 & PKS0405-123	& 0.16692		& 	$16.45\pm 0.05$		& $14.72\pm0.02$			& 5 & 3-7 \\
4 & PG1116+215	& 0.13847		& 	$16.20 \pm 0.04$		& $13.68^{+0.10}_{-0.08}$	& 2 & 8 \\
5 & PG1216+069	& 0.00632		& 	$19.32 \pm 0.03$		& $< 14.26$				& 2 & 9  \\
6 &	"	   	& 0.12360         &     	$> 15.95$               		&  $14.83 \pm 0.10$			& 9 & 10  \\
7 &   "                  & 0.28189         &       $16.70 \pm 0.04$		& $14.02 \pm 0.02$			& 3 & 10 \\
8& 3C 351.0		& 0.22111		& 	$> 17.0$ 				&  $14.27 \pm 0.04$			& 3 & 10 \\
9 &PHL1811		& 0.07765		& 	$16.03 \pm 0.07$		& $13.56 \pm 0.10$ 			& 2 & 10 \\
10 &	"			& 0.08093		& 	$17.98 \pm 0.05$		& $< 13.59$				& 2 & 11  \\
11 & J1009+0713 	& 0.3558   	&    $ 18.0 - 18.8$			& $15.0\pm0.2$			& 9 & 12 
\enddata
\tablerefs{
(1) \cite{2011arXiv1102.2850S}; 
(2) \cite{Lehner:09:734};
(3) \cite{Chen:00:L9};
(4) \cite{Prochaska:04:718}; 
(5) \cite{Williger:06:631};
(6) \cite{Lehner:07:680}; 
(7) \cite{Savage:10:1526}; 
(8) \cite{Sembach:04:351}; 
(9) \cite{Tripp:05:714}; 
(10) \cite{Tripp:08:39};
(11) \cite{Jenkins:05:767}.
(12) this paper.}
\label{llstable}
\end{deluxetable*}

The low ion \ion{Mg}{2} is another cold gas tracer that could provide information on the interface scenario. We have successfully built a photoionization scenario that works well for the detected \ion{Mg}{2} and other low ions. The simplest interface scenario suggests that each ``cold cloud'' could contribute two interfaces to the sightline. We have grouped the major kinematic components in this absorber into four ranges, but if components are generously defined to maximize their number for the \ion{Mg}{2} there could be approximately nine components over groups A - C, but none for group D. Thus at most 20 interfaces are expected in the \ion{O}{6}, not the $\gtrsim 100 $ that are implied by the observed column density and the interface models. While there could be numerous cold clouds and twice as many interfaces that fall below even the stringent detection limits of our HIRES data, there would need to be a large number of undetected \ion{Mg}{2} to provide enough interfaces to reproduce the \ion{O}{6}.

\begin{figure}[!t]
\epsscale{1.2}
\plotone{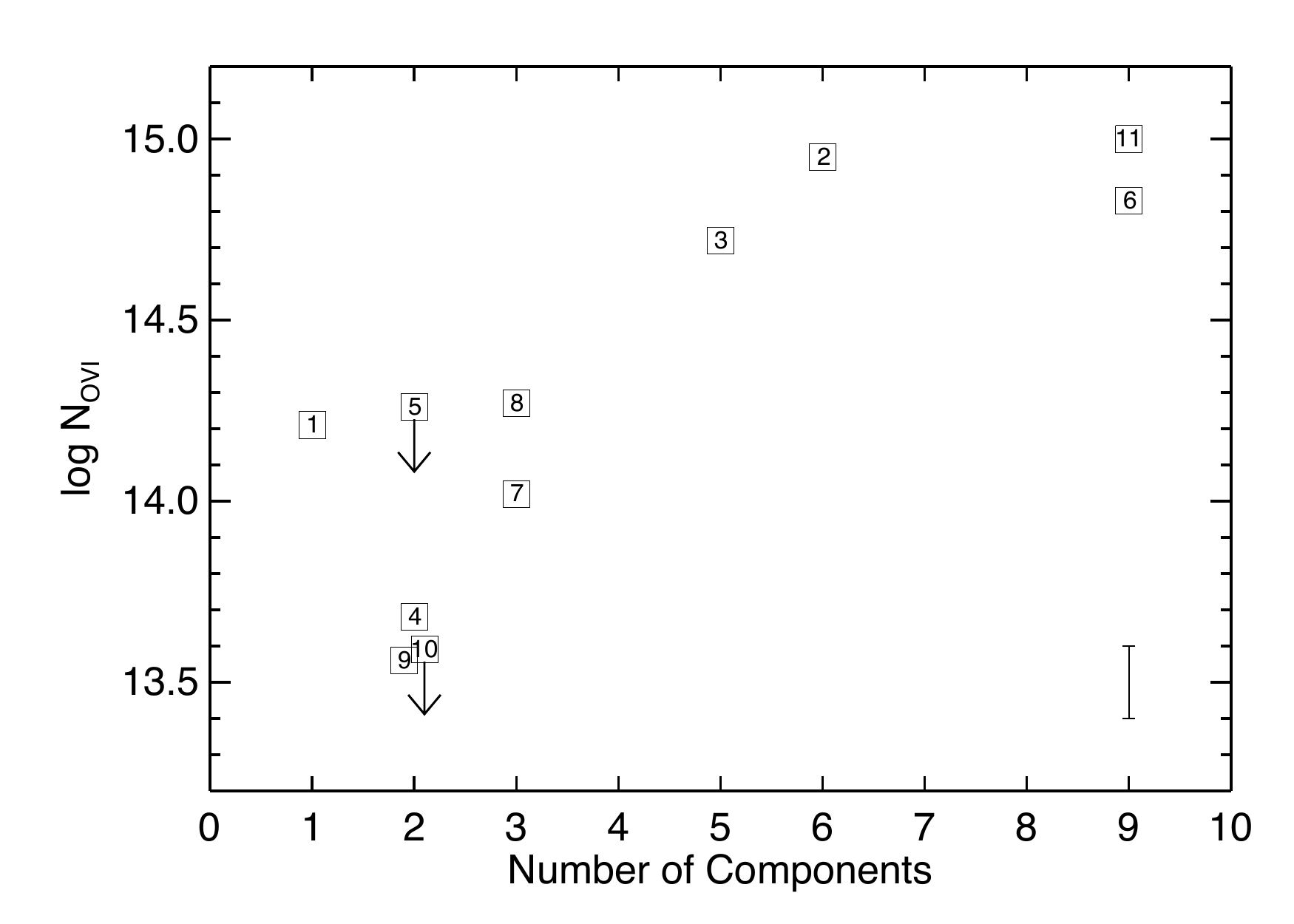}
\caption{ A correlation of the total system column density of \ion{O}{6} versus the number of reported kinematic components for the literature compilation reported in Table 3. The typical error bar is generously assigned to be $\pm0.1$ dex and is displayed at lower right; the apparent trend is well outside this margin of error. \label{o6compfig}}
\end{figure}

The features of this system recall the properties of some other low-redshift, \ion{O}{6}-bearing LLS that have been intensively studied to which we can compare and contrast our results. These systems are tabulated in Table 3, where we list all the systems in the literature with well-studied UV spectra covering \ion{O}{6} from {\it HST} or {\it FUSE} and having $\log N$(\ion{H}{1}) $> 16$ to within observational errors. The listed number of components is the number of \ion{H}{1} and/or metal-line components given in the references cited. Here we have assigned 9 components to the J1009+0713 LLS based on the number of observed \ion{Mg}{2} components in the HIRES data, which has comparable resolution to the STIS/E140M data used in most of the other cases from the literature.This compilation of results suggests two possible patterns of relevance to the J1009+0713 LLS and the origins of strong \ion{O}{6} absorbers. First, the total quantity of \ion{O}{6} is not well correlated with the column density of \ion{H}{1}. The former spans less than 1 order of magnitude while the latter covers more than three dex. This poor correlation is one of the central puzzles in the origins of the \ion{O}{6} absorbers (cf. Figure 6 of \citet{Thom:08:22}). Second, the total {\it system} column density of \ion{O}{6} appears to correlate with the number of detected components in the system, such that $> 4$ components gives $\log N$(\ion{O}{6}) $>14.5$, while systems with a smaller number of components have lower column density. This effect is clearly seen in Figure~\ref{o6compfig}, where we plot the \ion{O}{6} column density over the number of components from Table 3. 

While the sample is still small, there is a marked correlation that is well outside the typical error on individual measurements (conservatively assigned as $\pm 0.1$ dex and shown at lower right).  We apply a Spearman's rank correlation test and find that we can reject at 99.7\% confidence the null hypothesis that there is no correlation between the \ion{O}{6} column density for these systems and the number of kinematic components they exhibit\footnote{The correlation is still strong if we assign only the 4 components detected at COS resolution to the J1009+0713 LLS; the data then reject the null hypothesis at 99.3\% confidence.}. We emphasize that this is not a homogeneously selected sample, though we have tried to gather all reported systems that meet the selection criteria. Nevertheless we find at least a strong suggestion that a large number of components leads to a strong  \ion{O}{6} absorber. This correlation could indicate that the \ion{O}{6} is generally associated with the low-ionization material even if it does not reside in the same physical layers of the gas. This general behavior is expected in the interface scenario discussed above, though the problem of how just a few interfaces give $\log N$(\ion{O}{6}) $>14.5$ remains.

The J1009+0713 system and other systems like it thus offer mild but not conclusive evidence in favor of the interface scenario. The broadening of strong O VI into many detectable components and a correlation with a large number of velocity components in the \ion{H}{1} and low ions are consistent with the basic expectations of the interface scenario. But the total quantity of \ion{O}{6} substantially exceeds the expectations of the interface models, given the number of apparent cold gas clouds along the sightline. Three possible solutions to this puzzle are that a large number of cold clouds go undetected but their interfaces are seen, or the \ion{O}{6} from a single interface is well above that calculated in models, or the interface model does not apply. It is still possible, perhaps even likely, that some fraction of the \ion{O}{6} we observe is produced in interfaces, along with additional contributions from another mechanism. This system shows that further measurements of \ion{O}{6} and \ion{Mg}{2} can provide more information about the correlation of hot gas with low-ionization gas for the same sample of absorbers. Such a study is planned as part of our larger survey. 

\section{Summary and Discussion}
\label{interp-section}

We have examined the ionization, metallicity, and association with galaxies of a newly-discovered strong intervening \ion{O}{6} absorption-line system discovered in our COS data on J1009+0713. This system exhibits $\log N$(\ion{O}{6}) = 15.0 spread kinematically over 400 \kms. It appears to be associated with at least two galaxies at projected separations of 14 and 46 kpc from the sightline that have redshifts coincident with the detected gas. The system includes two similar LLS-strength components separated by 60-80 \kms\ with two weaker associated components 60 - 100 \kms\ away from these. The direct line measurements only constrain the gas metallicity in the two strongest components to $Z = 0.1 - 1 Z_{\odot}$ owing to saturation of the \ion{H}{1}. However,  the detected ion ratios are well-matched by a photoionization model in which simple clouds of total column density $\log N$(H) $\simeq 20$ and $5 - 50$\% solar metallicity are ionized to a 1 - 10\% neutral fraction by a radiation field from star forming regions at a distance of 20 - 100 kpc. In short, it appears that the bulk of this absorber traces gas that resides in the common halo environment of these galaxies and is ionized by their ongoing star formation. Similar modeling applied to the two outlying components, with $\log N$(\ion{H}{1}) = 14.8 and 16.5, are also consistent with their detected absorption for models with $\log U \sim -2$ and $\sim 25$\% solar metallicity. These outlying clouds also appear to be relatively highly ionized, with neutral fractions of 0.1 - 1\% even in the gas traced by low ions, but all with an additional, difficult-to-constrain, highly ionized component traced by \ion{O}{6}.

This system exhibits two galaxies ranging over two magnitudes in the rest-frame optical, the brightest (170\_9) has one with $M_* = 6 \times 10^9 M_{\odot}$ in stars\footnote{The galaxy 92\_7 appears in the SDSS photometric database with $r = 22$, and appears distinctly in the WFC3 image, but we have not yet obtained its redshift. It may also be associated with this absorption-line system.}. Galaxy 86\_4 is 2 mag fainter with a total star formation rate of only about 1/10 that of 170\_9. Based on emission-line measurements and its inferred stellar mass, the brighter galaxy 170\_9 appears to be consistent with solar metallicity, while 86\_4 is either consistent with solar metallicity or with $\sim 25$\% solar (depending on which branch of the R23 indicator is adopted). Since none of the detected gas components appear to {\it require} a solar metallicity (though B and C are consistent with solar if the photoionization modeling is discounted), we do not have positive evidence that any of the detected gas was brought into the sightline by outflows from the galaxies.


Several features of specific systems are worth comparing to the J1009+0713 LLS. The system at $z = 0.167$ toward PKS0405-1219 \citep{Chen:00:L9}  lies $\simeq 70$ kpc from two massive galaxies, is at least 0.1 solar and likely higher metallicity, and exhibits strong absorption from high ions up to \ion{N}{5} and \ion{O}{6} at $\log N$(\ion{O}{6}) = 14.6. The $z = 0.0809$ LLS toward PHL 1811 \citep{Jenkins:05:767} lies 34 and 87 kpc from two nearby L* galaxies, and has nearly a solar metallicity, but is not associated with highly ionized gas. Both of these absorbers show evidence of a predominantly ionized gas, though the PHL 1811 system also exhibits absorption by low ions. While our simple photoionization models shown above indicate a sub-solar metallicity for the J1009+0713 system, this model-dependent value may not reflect the true gas metallicity in the LLS components: the \ion{O}{1} measurements are consistent with solar metallicity.  The PKS0405-1219 system is a particularly interesting comparison to the J1009+0713 system, since the \ion{O}{6} in both cases is stronger and more kinematically complex than would be expected from the lower ionization species. While is it difficult to draw general conclusions from only three cases, it appears that galaxy halos may commonly host LLS absorption, that it can be quite enriched in metals to nearly solar abundances, and that it can be associated with large quantities of highly ionized gas. Based on the PKS0405-1219 and J1009+0713 systems, we speculate that LLS-strength absorption may be necessary to produce the strongest intervening \ion{O}{6} absorbers, $\log N$(\ion{O}{6}) $\gtrsim 14.5$. The general incidence of LLS in galaxy halos and its relationship to hot gas will be addressed by our larger survey, which should also be able to assess the gas metallicity in many cases. High metallicity in dense, cool gas located at $> 50$ kpc from the nearest galaxies may indicate a major role for tidal stripping of dwarf satellites, or robust outflows from large galaxies or their dwarf satellites. 

This sightline passes by a galaxy with nearly $L*$ and a dwarf galaxy possibly in its halo or interacting with it, which may have low metallicity. Their total velocity separation is only $25$ km s$^{-1}$, which suggests they are gravitationally associated. It is tempting to draw comparisons of this system to the Milky Way and its Magellanic Clouds. If the smaller galaxy has the metallicity of $25$\% indicated by their lower-branch R23 measurements, it could be associated with the A and/or D components in an outflow or tidal-stripping scenario. If so, some of the detected absorption components may arise in gas stripped from this dwarf galaxies, like the Magellanic Stream.

Alternatively, all the detected absorption components could have sub-solar metallicity and could trace low-metallicity, infalling material not unlike the large HVC complexes of the Milky Way. This inference and the high ionization state of the detected gas would imply that infalling, metal-poor gas enters halos and/or galaxies in ionized form, and is possibly associated with still more highly ionized material traced by \ion{O}{6} that arises either in the hot diffuse galaxy halo itself or in interface zones of intermediate temperature between that hot ambient medium and the cooler infalling clouds. This scenario has been well developed for the Milky Way - further tests of its validity for external galaxies requires better knowledge of gas metallicities, kinematics in association to galaxies, and thus more well-studied cases.  

In our combined COS and HIRES data we have used two important diagnostic lines for gas in galaxy halos -- both \ion{O}{6} and \ion{Mg}{2} -- that are typically not available in combination for the practical reason that they require both UV and optical data and galaxies at $z \gtrsim 0.1$. Recently, \cite{Barton:09:1817}  and \cite{Chen:10:1521} have systematically addressed the number density of \ion{Mg}{2} absorbers in the $\sim 100$ kpc regions surrounding galaxies using galaxies and QSO sightlines selected from SDSS. They find that the equivalent width of \ion{Mg}{2} increases at low impact parameter in a well-defined trend that has a detectable dependence on galaxy luminosity for those galaxies that exhibit \ion{Mg}{2} absorption. However, both \cite{Bowen:11:47} and \cite{Gauthier:10:1263} have found lower covering fractions of strong \ion{Mg}{2} surrounding more massive galaxies (LRGs) that indicate the cool gas covering fraction depends, perhaps strongly, on galaxy type and/or environment. For comparison with these samples, we obtained a total $W_r^{2796} = 1250 \pm 15$ m\AA\ for the full LLS analyzed here, including 188 m\AA\ in component group A, 604 m\AA\ in component group B, and 472 m\AA\ in component group C. These findings are all within the range of the distributions found by these other studies. However, since we also detect \ion{H}{1}, \ion{O}{6} and other multiphase UV ions in these component groups, we can draw additional implications that are not possible based on \ion{Mg}{2} alone. 

First, we have used the detections of \ion{Mg}{2} and the ionization stages of C, N, O, Mg, S, Si, and Fe to infer that the absorbing clouds have only a small portion, 0.1 - 1\% of their gas in the neutral phase. This finding implies that the gas mass traced by these absorbers could be significant. Indeed, \cite{Chen:10:1521} inferred \ion{Mg}{2} cloud masses of $\sim 2 \times 10^4$ M$_{\odot}$ and a total baryonic mass in halos of $3 \times 10^9$ M$_{\odot}$, using the important assumptions of 10\% solar metallicity and a 10\% ionization fraction for \ion{Mg}{2}. These values are within the range permitted for the J1009+0713 clouds, where the ionization corrections and metallicities are better constrained by multiple ionization stages. Thus our findings indicate that the assumption of a significant mass correction to the observed \ion{Mg}{2} is reasonable, and such absorbers could harbor significant mass. Moreover, the \ion{O}{6} suggests a possibly significant contribution of mass over and above that provided by \ion{Mg}{2} systems. Finally, it may prove important to measure halo absorbers in {\em both} \ion{O}{6} and \ion{Mg}{2}; components A, B, and C here hint that \ion{O}{6} and \ion{Mg}{2} can have some relationship but may vary in their ratio by over an order of magnitude. Our larger survey was designed with the goal of detecting both lines in a significant sample of galaxies over a range of mass. We will thus be able to test the findings of  \cite{Barton:09:1817}  and \cite{Chen:10:1521} that \ion{Mg}{2} correlates with galaxy luminosity, and also the finding by \cite{Bowen:11:47} and \cite{Gauthier:10:1263} that massive galaxies have less \ion{Mg}{2}, while also exploring these relations in \ion{O}{6} for the same galaxies. These previous results, considered together with hints of a \ion{Mg}{2} / \ion{O}{6} relationship for the J1009+0713 system, suggests this will be a fruitful line of research. 

While it is among the strongest known \6\ absorbers, this system at least superficially resembles the highly-ionized, multiphase, \6-bearing HVCs surrounding the Millky Way. For example, \cite{Fox:05:332} report conditions very similar to those we find here for two HVC complexes toward HE0226-4110 and PG0953+414 in the Milky Way halo. Specifically, their ionization modeling finds $\log U \simeq -3.5$, subsolar metallicity, and $\log n_H \simeq -2$, but for clouds with $\log N$(\ion{H}{1}) $\sim 16-16.5$. Like them, we have found consistent photoionization models for the present system that yet do not explain the observed abundance of \6. Our findings also resonate with the recent survey of ionized silicon in the Milky Way HVC population by \cite{Shull:09:754} and \cite{Collins:09:962}, who find typical values of $\log U \sim -3$ and 10 - 30\% solar metallicity or lower in these clouds with neutral fractions of $\simeq 1$\%. These favorable comparisons suggests that highly-ionized, possibly metal-poor gas resides in the halos of star forming galaxies as it does in the Milky Way -- a key motivation for the survey that discovered this absorber. A more speculative but interesting inference can be drawn from the column density and ionization of the main components B and C. Were they exposed to a lower radiation field, because of a lower galactic $f_{esc}$ or SFR, they could instead appear as a sub-DLA or DLA with $N$(\ion{H}{1}) $\sim 10^{20}$ cm$^{-2}$ located at least 20 kpc from the nearest galaxy, and might have total column densities similar to the classical MW HVC complexes. We note that recent surveys for 21-cm absorption affiliated with nearby galaxies \citep{Borthakur:10:131, Gupta:10:849} have found 21-cm absorbers to be quite rare in galaxy halos.  While this could be an indication that sub-DLAs and DLAs are rarely found at such large impact parameters \citep[cf.][]{Meiring:11:2516, Peroux:11:2251}, this remains an open question since 21-cm absorption depends on spin temperature, and sub-DLAs and DLAs might be missed in 21-cm absorption surveys if the gas is too warm.  Our larger survey will address this problem by assessing the incidence of LLS, sub-DLAs, and DLAs in galaxy halos.

While the similarity of this absorber to some well-studied Milky Way HVCs provides an encouraging link between the HVCs and galaxy halo gas at large, this system also illustrates some important limitations that must be overcome to fully map out diffuse gas in halos. First, \ion{H}{1} column densities in the range that gives an LLS but not a DLA, as here, are extremely difficult to constrain to a narrow range. This is especially true if the gas obviously occupies several clearly separated kinematic components that are blended in the Lyman series, as shown by our HIRES data. This unfortunate circumstance makes metallicities hard to estimate, even in cases where conditions are otherwise favorable for obtaining metal-line column densities, as they are here. While LLS are rare in the IGM at large, they may be quite common close to galaxies where total gas densities are high, even if the gas is typically highly ionized. This system also illustrates vividly that relating halo gas to its host galaxies is complicated by projection effects and by multiple galaxies lying along the sightline. We clearly cannot associate the absorption, especially in the \ion{O}{6}, with any of the three galaxy candidates in particular. 

Finally, this system illustrates that detection of many ionization stages can constrain the physical parameters of the gas, but only for well-defined hypothetical physical scenarios, e.g. pure photoionization or pure collisional ionization. Combinations of such models, or non-standard scenarios such as interfaces and/or non-equilibrium calculations rapidly become too complex to test easily. Even for simple scenarios it is sometimes difficult to identify unique explanations for detected absorption. These considerations suggest that a large, systematic survey to assess possible variations could help  by smoothing out variations along individual lines of sight. Our systematic survey of gas in galaxy halos, which serendipitously discovered this absorber, aims to establish a set of empirical facts about the content of gaseous halos. Yet multiphase, kinematically complex absorbers like this one pose a stiff challenge to our ability to build theoretical models that are up to the task of relating complex physical processes in galaxy halos to observable quantities. Progress in this area could significantly advance the cause of understanding galaxy formation at large. 

\acknowledgments

We are happy to acknowledge a very constructive referee's report from Mike Shull, and helpful comments from Andrew Fox. Support for program GO11598 was provided by NASA through a grant from the Space Telescope Science Institute, which is operated by the Association of Universities for Research in Astronomy, Inc., under NASA contract NAS 5-26555. TMT appreciates support for this work from NASA ADP grant NNX08AJ44G. Some of the data presented herein were obtained at the W.M. Keck Observatory, which is operated as a scientific partnership among the California Institute of Technology, the University of California and the National Aeronautics and Space Administration. The Observatory was made possible by the generous financial support of the W.M. Keck Foundation. The authors wish to recognize and acknowledge the very significant cultural role and reverence that the summit of Mauna Kea has always had within the indigenous Hawaiian community.  We are most fortunate to have the opportunity to conduct observations from this mountain.

{\it Facilities:} \facility{HST (COS, WFC3)}, \facility{Keck (LRIS, HIRES)}.


\bibliographystyle{/Users/tumlinson/astronat/apj/apj}

\bibliography{test,test2}

\end{document}

%% file: t1.tex
\begin{deluxetable*}{lcccc}
\vspace{-0.3in}
\tablecolumns{6} 
\tablenum{1} 
\tablewidth{0pt} 
\tablecaption{Measured Component Column Densities in the $z = 0.3558$ System} 
\tablehead{\colhead{Line}
&\colhead{A}
&\colhead{B} 
&\colhead{C}
&\colhead{D}\\  
\colhead{Centroid} & \colhead{$-95$ \kms} & \colhead{$-10$ \kms} & \colhead{$+60$ \kms} & \colhead{$+180$ \kms}  \\
\colhead{Range} & \colhead{$-200$ to $-65$} & \colhead{$-65$ to $30$} & \colhead{$30$ to $130$} & \colhead{$130$ to $250$} 
} 
\startdata 
Ly$\alpha$ 1215		& blend						&  \multicolumn{2}{c}{blend}					& blend					 \\
Ly$\beta$ 1025 		& blend						& \multicolumn{2}{c}{blend}					& 14.83,+183,17\tnote{a}		 \\
Ly$\gamma$ 972		& blend						& \multicolumn{2}{c}{blend}								& 14.79,+180,16\tnote{a}		 \\
Ly$\delta$ 949			& blend						&  \multicolumn{2}{c}{blend}								& contaminated			 \\
Ly$\epsilon$ 937		& blend						&  \multicolumn{2}{c}{blend}								& contaminated			 \\
Ly$\zeta$ 930  			& blend						&  \multicolumn{2}{c}{blend}								& \nodata					 \\
Ly$\eta$ 926 			& 16.49,-95,17\tablenotemark{a}	& \multicolumn{2}{c}{blend}						& \nodata	\\
Ly$\theta$ 923 			& 16.33,-93,22\tablenotemark{a}	&  \multicolumn{2}{c}{blend}								& \nodata	\\
Ly$\iota$ 920 			& contaminated				& \multicolumn{2}{c}{blend}						& \nodata	\\
Ly$\kappa$ 919 		& 16.44,-94,20\tablenotemark{a} 	&  \multicolumn{2}{c}{blend}								& \nodata	\\
Ly$\lambda$ 918 		& not fitted\tablenotemark{a}		& \multicolumn{2}{c}{blend}								& \nodata	\\
Ly$\mu$ 917			& blend						& \multicolumn{2}{c}{blend}								& \nodata	\\

					& & & & \\

C I 1157   				& $<14.4$						& $<14.3$					& $<14.3$            				& $<14.4$  				 \\ 			
C II 1036\tnote{e} 				
 				         & $14.0\pm0.1$ 				& $14.6\pm0.3$ 			& $14.6\pm0.3$  	   			& $<13.5$ 					\\ 
C III 977\tnote{f}		
					& $>14.3 \pm 0.2$				& $>14.3 \pm 0.2$                         & $>14.2 \pm 0.2$                                   & $>13.5 \pm 0.1$                         \\
					
					& & & & \\
					
N I 1199, 1200			& $<13.7$						& $<13.7$ 				& $<13.7$            				&$<13.7 $					 \\
N II 1083				& $<13.4$						& 14.1 $\pm$ 0.1			& $14.0 \pm 0.1$                  		& $<13.6$ 				 \\ 
N III 989\tnote{b}		& $14.3 \pm 0.1$				& $14.6 \pm 0.1$			& $14.5 \pm 0.1$ 				& $14.5 \pm 0.1$	 		 \\
N V 1238,1242			& $<13.6$						& $<13.7$ 				& $<13.7$            				& $<13.7$					 \\ 

					& & & & \\

O I 1302				& $<13.9$						& $14.6\pm 0.2$			& $14.1\pm0.2$	                		& $<13.9$ 				 \\ 
O I 988				& $<14.0$						& $14.4\pm 0.2$			& $14.2\pm0.2$                 			& \nodata\tnote{c}	 		 \\
OVI 1032				& $14.67\pm0.1$				& $14.40\pm0.1$			& $14.40\pm0.1$     				& $14.29\pm0.1$			 \\
OVI 1038				& $14.56\pm0.1$				& $14.38\pm0.1$			& $14.26\pm0.1$ 				& $14.33\pm0.1$			 \\
					
					& & & & \\

Mg I 2852				& $<11.2$						& $11.9 \pm 0.04$			& $11.4\pm 0.1$				& $<11.2$ 				  \\ 
Mg II 2796\tnote{e}		& $12.82\pm0.02$				& $13.59\pm0.04$			& $13.45\pm0.04$          			& $<11.2$ 				  \\ 
Mg II 2803\tnote{e}		& $12.99\pm0.02$				& $13.75\pm0.03$			& $13.62\pm0.03$          			& $<11.5$ 				  \\ 
					
					& & & & \\
					
Si I  2515				& $<11.7$						& $<11.7$					& $<11.7$   					& $<11.7$ 				 \\ 
Si II 1260\tnote{f}		& $>12.7\pm0.3$	                            & $>13.6\pm0.2$      			& $>13.6\pm0.2$  				&$<12.7$					 \\
Si II 1190		               	& \nodata\tnote{c}                          	& $13.9\pm0.2$       			& $13.8\pm 0.2$  				&  $<13.6$            			  \\
Si II 1193				& \nodata\tnote{d}                              	& $13.8\pm0.2$      			& $13.5\pm0.2$    				& $<13.4$ 				 \\
Si III 1206\tnote{f}		
					& $>13.5\pm0.2$ 				& $>13.8\pm0.2$  			& $>13.8\pm0.2$                    		& $>12.6\pm0.2$ 	 		 \\
	
					& & & & \\

S II 1253\tnote{g}      			& $<14.8$						& $<14.7$					& $<14.7$ 					& $<14.7$ 				  \\ 
S III 1012				& $14.2 \pm 0.1$				& $14.3 \pm 0.1$			&$13.8 \pm 0.1$				& $<13.4$ 				 \\
S IV 1062				& $14.0 \pm 0.1$				& $<13.8$					& $<13.8 $   					& $<13.8$ 				 \\ 

					& & & & \\

Ca II 3934				& $<11.4$						& $11.50 \pm 0.15$ 			& $<11.4$	   					& $<11.4$	 				  \\ 
Ca II 3968				& $<11.9$						& $<11.9$					& $<11.9$   					& $<11.9$ 				 \\ 

					& & & & \\

Ti II 3384				& $<11.5$						& $<11.5$					& $<11.5$   					& $<11.5$ 				 \\ 

					& & & & \\

Fe I 2484				& $<11.3$						& $<11.3$					& $<11.3$   					& $<11.3$ 				 \\ 
Fe II 2382                            	& $<11.8$        					& $13.71 \pm 0.4$       		& $13.45 \pm 0.16$          			&  $<11.8$               			 \\
Fe II 2600 			& $<11.6$						& $13.65 \pm 0.07$			& $13.37\pm 0.02$          			& $<11.6$ 				  \\ 
Fe II 2586				& $<12.2$						& $13.88 \pm 0.03$			& $13.38\pm 0.04$			         	& $<12.2$ 				 \\ 
Fe II 1144				& $<13.8$						& $14.1\pm0.2$			&  \nodata\tnote{c}          			&  \nodata\tnote{c} 			 \\ 
Fe II 2374				& $<12.8$						& $14.02 \pm 0.09$ 			&  $<12.8$   					&  $<12.8$ 				 \\ 
Fe II 1063				& $<13.8$						& $13.7 \pm 0.2$			& $13.6 \pm 0.2$   				& $<13.8$ 				 \\ 
Fe III 1122			& \nodata\tnote{c}				& $14.3\pm 0.1$			& $14.2\pm 0.1$          			& $<13.9$	 				 
\enddata 
\label{comp-list}
\tablenotetext{a}{Part of the definition of this component;  HI where possible, metal lines where necessary.}
\tablenotetext{b}{N III $\lambda$989.80 column densities are uncertain because they are blended with \ion{Si}{2} $\lambda$989.87 to an unknown degree. These column densities should be treated as upper limits.  Component D is even more uncertain, because it appears to be blended with other unidentified absorption.}
\tablenotetext{c}{Contaminated by unknown absorption - no measurement or limit is possible.}
\tablenotetext{d}{Very noisy.}
\tablenotetext{e}{Subject to mild saturation.}
\tablenotetext{f}{Strongly saturated and/or line black absorption. The equivalent widths over each velocity range are converted to column densities assuming a linear curve of growth, which yield robust lower limits.}
\tablenotetext{g}{\ion{S}{2} $\lambda$1259 is contaminated and $\lambda$1253 and $\lambda$1250 are undetected. The $\lambda$1253 line provides the best upper limits. } 
\tablecomments{All column density measurements were obtained via direct integration of the line profiles over the given velocity ranges, except where noted in the comments field. Upper limits for UV lines are obtained as $2\sigma$ limits by direct integration of the apparent column densities over the stated component velocity ranges. For multiplets, the limits use the strongest uncontaminated transition. Line profile fitting results are given as a triplet; column density, velocity centroid, doppler b-value.}
\end{deluxetable*} 